\documentclass[%
reprint,
superscriptaddress,
notitlepage,
nofootinbib,
 amsmath,amssymb,amsfonts
floatfix,
]{revtex4-1}
\usepackage{mathtools}
\usepackage{amsthm}
\usepackage{appendix}
\usepackage{physics}
\usepackage{graphicx}
\usepackage{dcolumn}
\usepackage{bm}
\usepackage{hyperref}
\usepackage[normalem]{ulem}
\usepackage{xcolor}

\begin{document}

\preprint{}

\title{Bath assisted transport in a three-site spin chain:\\ global {\sl vs} local approach}

\author{F. Benatti}
\affiliation{Department of Physics, University of Trieste, I-34151 Trieste, Italy}
\affiliation{Istituto Nazionale di Fisica Nucleare (INFN), Sezione di Trieste, I-34151 Trieste, Italy}
\author{R. Floreanini}
\affiliation{Istituto Nazionale di Fisica Nucleare (INFN), Sezione di Trieste, I-34151 Trieste, Italy}
\email{f@infn}
\author{L. Memarzadeh}%
\email{memarzadeh@sharif.edu}
\affiliation{Department of Physics, Sharif University of Technology, Tehran, Iran}%


\begin{abstract}

Within the standard weak-coupling limit, the reduced dynamics of open quantum spin chains with their two end spins coupled
to two distinct heat baths at different temperatures are mainly derived using the so-called
\textit{global} and \textit{local} approaches, in which, respectively, the spin self-interaction
is and is not taken into account. In order to compare the differences between the two regimes, we concentrate on an open three-site $XX$ spin-chain, 
provide systematic techniques to addressing the global and local asymptotic states and then compare the asymptotic spin-transport features by studying the spin flux
through the middle site. 
Basing on the analytical expressions of the stationary states in the two regimes, we find that the local approach  misses important global effects emerging as spin sink and source terms that can only be due to non negligible inter-spin interactions. Moreover, we show that the local asympotic transport features cannot be recovered from the global ones by letting the inter-spin coupling vanish thus pointing to the existence of different coupling ranges where only one approach is physically tenable and possibly a region where the two descriptions may coexist.
\end{abstract}

\maketitle


\section{Introduction}
\label{sec:introduction}
Transport phenomena at the quantum scale have recently been receiving increasing attention,
as they are of fundamental importance both in theory, for understanding the behaviour of driven many-body systems,
and in applications, for the development of new quantum devices. The paradigmatic models for such studies are
provided by linear chains of spins, coupled among themselves and further interacting with external baths through the spins at their ends;
they allow modelling various instances of spin currents, possibly with controlled flux manipulation.
Indeed, many results on the dynamics of specific realizations of such systems have been reported in the recent 
literature with direct applications to ultracold-atoms, light-harvesting complexes and quantum thermodynamics at large.%
\footnote{The literature on the topic is vast; for instance, see \cite{Datta}-\cite{Prosen5}}

In the presence of external baths, any quantum system needs to be treated as ``open'', and its reduced dynamics,
obtained by tracing over the baths degrees of freedom, becomes non unitary.
In the so-called weak-coupling limit, in which the strength of the system-baths interaction is small,
the system time evolution can be conveniently described in terms of a master equation
in Gorini-Kossakowski-Sudarshan-Lindblad form, encoding effects of decoherence and dissipation \cite{Alicki-Lendi}-\cite{Merkli}.

For a spin chain, {\it i.e.} for a system made of many interacting subsystems, the derivation of such
master equation might be problematic. Indeed, due to the coupling among the spins,
a so-called {\it global} master equation should emerge, that requires the diagonalization
of the starting spin-chain Hamiltonian to be spelled out. The resulting dissipative dynamics  is expected to favour
environment induced excitation transfer between different sites ({\it e.g.} see \cite{Davies4}-\cite{Rivas2}).
However, its explicit derivation could be quite difficult. 

For these reasons, an alternative approach has been often followed for sufficiently small inter-spin couplings;
this leads to a so-called {\it local} master equation; indeed, in its derivation the spin-spin interactions are neglected
and only the local couplings of the spins at the two chain ends with the baths are taken into account ({\it e.g.} see \cite{Michel}-\cite{Hovhannisyan}). 
As a result, in this approach the decoherence and
dissipative effects involve only the spins directly coupled to the external baths.

A stream of different investigations ensued with the purpose of comparing the virtues and weaknesses of
the two point of views \cite{Rivas1,Guimaraes,Werlang,Santos,Migliore,Zoubi}, \cite{Levy}-\cite{Cattaneo}. 
The debate is still unsettled and both alternatives are regularly adopted in applications.

Aim of the present investigation is to contribute to the ongoing debate by 
an analytic investigation of the time-asymptotic features of a typical model
of quantum transport: a spin-1/2 chain, with $XX$-type interaction, 
in the presence of a constant transverse magnetic field, weakly coupled by means of its two end spins to two
separate heat baths at different temperatures. In order to be able to obtain a completely analytic description
of the chain reduced dynamics, we shall limit the discussion to a chain formed by just three sites and focus on the system transport properties 
corresponding to the rate of change in time of the average of the spin along the $z$ direction at the middle site. 
We derive the exact stationary state in the global approach and apply a systematic method to finding the stationary state in the local approach up to the first order perturbation expansion with respect to the inter-spin interaction. The analytic expressions obtained allows us to compare the asymptotic spin-transport properties in the two regimes without the ad hoc assumptions necessarily adopted in numerical studies.\\
In particular, we will show that the global approach corresponds to a physical regime where, beside the currents,  spin sink and source terms appear that are not present within the local approach. Indeed, we will see that these novel contributions to the spin continuity equation appear only because the Lindblad operators in the 
master equation derived in the global approach involve all three spins, while in the local approach the Lindblad operators pertain only to the leftmost and rightmost spins, those directly coupled to the baths.
Moreover, we shall also see that the structure of the steady states in the two regimes makes the local approach not recoverable from the global one in the limit of vanishing inter-spin interaction. However, when seen from the point of view of the sink and source contributions, such a discontinuity is small and becomes less and less visible with decreasing temperature difference between the baths. The discontinuity reflects  the lack of interchangeability between the ergodic average utilized in the derivation of the Lindblad master equation in the global approach and the switching off  the inter-spin interaction. Indeed, for small couplings the spin transition 
frequencies are close to degeneracy and the weak-coupling limit techniques in the global approach fail. On the contrary, away from degeneracy, when the inter-spin couplings become of the order of the transverse magnetic field, the sink and source terms clearly discriminate between the global and local approaches.  \\
With respect to the ongoing debate about the two approaches, these results indicate that, for sufficiently weak spin interactions and sufficiently high temperature, the local approach is the only one valid, while, for sufficiently strong couplings at any given temperature, the global approach is the only tenable one, with probably a range of couplings where the global approach blends with the local one.\\
In what follows, we focus upon the asymptotic properties of the open chain and not on the different scales and features characterizing the transient dynamics. Yet, the explicit analytic form of 
the stationary states, their remarkably different physical  features and the methods employed for their derivation may allow for analytic extensions to 
larger spin chains. In addition,  they may foster numerical investigations of the ranges of validity of the local and global approaches and of the parameter regions where they might coexist. Not to mention the possibility of the experimental verification of the presence of asymptotic sink and source terms, or, as discussed in the final section, the different transient features expounded by the currents in the two approaches, that would certainly discriminate between the feasibility of the global versus the local approach.
\\
The structure of the paper is as follows: in section II
we shortly review the standard weak-coupling limit background for deriving master equations of Gorini-Kossakowski-Sudarshan-Lindblad type. In Sections III and IV  
we obtain the master equations in the global, respectively the local approach, we compute the stationary states and analyze the corresponding asymptotic transport properties. 
In section V we conclude by summarizing and discussing the results, while the more technical issues are presented in the Appendices.

\section{Open $XX$ spin chain}
\label{sec:system}
As mentioned above, purpose of this work is the analytical study of the asymptotic transport properties of open quantum spin chains interacting with two thermal 
baths coupled to their end spins in 
the so-called global and local approaches; in order to achieve our goal, we restrict to  the simplest setting of a three-site spin-1/2  chain, whereby the 
steady states of the open reduced dynamics can be analytically accessed in both regimes and the corresponding transport properties addressed by looking 
at the middle spin. In this section we shortly review the necessary techniques that will subsequently be applied to extract from the closed dynamics of 
the spin chain interacting with the thermal baths a fully physically consistent reduced Markovian master equation for the three spins of the chain alone.

The closed spin dynamics will be given by a  nearest-neighbour $XX$-type inter-spin interaction 
in the presence of a transverse constant magnetic field of strength $\Delta$, with Hamiltonian:
\begin{equation}
H_S=g\sum_{i=1}^2\left(\sigma_{x}^{(i)}\sigma_{x}^{(i+1)}+\sigma_{y}^{(i)}\sigma_{y}^{(i+1)}\right)+
\Delta\sum_{i=1}^3\sigma_z^{(i)}\ ,
\label{spin-hamiltonian}
\end{equation}
where $\sigma_{x,y,z}^{(i)}$ are Pauli matrices attached to site $i$, and $g$ is the spin coupling constant;
in absence of the inter-spin interaction, the magnetic field contribution 
plays the role of a `free' Hamiltonian.

\begin{figure}
    \centering
    \includegraphics[width=1\linewidth]{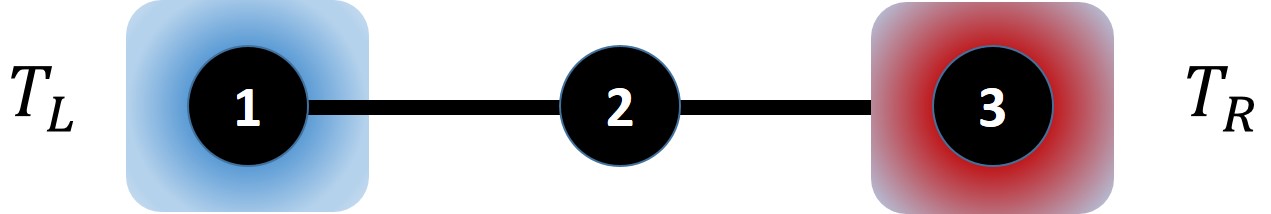}
    \caption{Three-spin chain in a two bath environment: first spin coupled  to the left bath at temperature
$T_L$  and third spin coupled to the right bath at temperature $T_R$.}
    \label{Fig1}
\end{figure}

We then turn the spin chain into an open quantum system by coupling the two external spins to two independent Bosonic thermal baths (see Fig.~\ref{Fig1}). 
We shall describe them  by two sets of independent 
mode operators, $b_\alpha(\nu)$, $b_\alpha^\dagger(\nu)$,
labelled by the discrete index $\alpha=L,R$, distinguishing the two baths, and by the continuous
variable $\nu$, obeying standard commutation relations, 
$[b_\alpha(\nu),\, b_\beta^\dag(\nu')]=\delta_{\alpha\beta}\, \delta(\nu-\nu')$.
From them it follows that in natural units where both Planck and Boltzmann constants are set to $1$, $\hbar=\kappa_B=1$, the operators $b_\alpha(\nu)$and $b_\alpha^\dagger(\nu)$ have dimension $E^{-1/2}$, where $E$ stands for energy.

Despite their infinitely many degrees of freedom, for sake of simplicity we shall denote by  $H_B^{(L)}$ and $H_B^{(R)}$ their free Hamiltonians 
and by 
\begin{align}
\label{freedyn1}
{\rm e}^{itH_B}\,b_\alpha(\nu)\,{\rm e}^{-itH_B}&={\rm e}^{-i \nu t} \, b_\alpha(\nu)\\ 
\label{freedyn2}
{\rm e}^{itH_B}\,b^\dag_\alpha(\nu)\,{\rm e}^{-itH_B}&={\rm e}^{i \nu t} \, b_\alpha(\nu)
\end{align}
their free dynamics with $H_B=H_B^{(L)}+H_B^{(R)}$.

The coupling of the baths to the spin chain, the $L$-bath to the first spin, the $R$-bath to the third one,
is supposed to be weak and described by a typical system-environment Hamiltonian $H'$ of the form:
\begin{equation}
H'=\sum_{\alpha=L,R} \Big( \sigma_+^{(\alpha)} B_\alpha + \sigma_-^{(\alpha)} B_\alpha^\dag \Big)\ ,
\label{interaction-h}
\end{equation}
where 
\begin{equation}
\sigma_\pm^{(L)} \equiv \frac{1}{2}\big( \sigma^{(1)}_x\pm i \sigma^{(1)} _y\big)\ ,\quad
\sigma_\pm^{(R)} \equiv \frac{1}{2}\big( \sigma^{(3)}_x\pm i \sigma^{(3)}_y \big)\ ,
\label{sigmas}
\end{equation}
are spin variables of the first and the third site, while
\begin{equation}
B_\alpha=\int_0^\infty {\rm d}\nu\, h_\alpha(\nu)\, b_\alpha(\nu)\ ,\quad [h_\alpha(\nu)]^* = h_\alpha(\nu)\ ,
\label{bath-operator}
\end{equation}
are the corresponding bath operators, where $*$ means complex conjugation. 
Notice that the role of the real functions $h_\alpha(\nu)$ is that of smearing functions  introducing
an effective cutoff in the above $\nu$ integrals in order to make the bath operators $B_\alpha$ well-defined.%
Furthermore, the smearing functions will be taken to have dimension $E^{1/2}$, so that the dimension of 
the operators $B_\alpha$ and $B_\alpha^\dag$ is $E$.
The total Hamiltonian $H$ describing the complete system, the spin-chain
together with the two external baths, can thus be written as
\begin{equation}
\label{totHam}
H=H_0 + \lambda\, H' \quad \hbox{where}\quad H_0=H_S +H_B\ ,
\end{equation}
with $\lambda\ll1$ a small dimensionless coupling constant.
The Hamiltonian $H$  generates the evolution in time of the 
total density matrix $\rho_{\rm tot}$,
$\partial_t\rho_{\rm tot}(t)=-i[H,\ \rho_{\rm tot}(t)]$,
starting at $t=\,0$ from the initial
total state  $\rho_{\rm tot}(0)$.
We shall assume chain and baths to be initially prepared 
in an uncorrelated state, with the statistically independent thermal baths in
their equilibrium Gibbs states, whence $\rho_\beta=\rho_{\beta_L}\otimes\rho_{\beta_R}$, 
characterized by temperatures $T_L\equiv1/\beta_L$ and
$T_R\equiv1/\beta_R$, respectively. Namely, 
\begin{equation}
\label{Gibbs}
\rho_\beta=\frac{{\rm e}^{-\beta_L\,H_B^{(L)}}}{{\rm Tr}\Big({\rm e}^{-\beta_L\,H_B^{(L)}}\Big)}\,\otimes\,
\frac{{\rm e}^{-\beta_R\,H_B^{(R})}}{{\rm Tr}\Big({\rm e}^{-\beta_R\,H_B^{(R)}}\Big)}\ ,
\end{equation}
whence the thermal expectations
\begin{align}
\label{expectation1}
&\hskip-.2cm
{\rm Tr}_B\Big(\rho_\beta\,b^\dag_\alpha(\nu)\,b_{\alpha'}(\nu')\Big)=\delta_{\alpha\alpha'}\delta(\nu-\nu')\,n_\alpha(\nu)\\
\label{expectation2}
&\hskip-.2cm
{\rm Tr}_B\Big(\rho_\beta\,b_\alpha(\nu)\,b^\dag_{\alpha'}(\nu')\Big)=\delta_{\alpha\alpha'}\delta(\nu-\nu')\,(1+n_\alpha(\nu))\ ,
\end{align}
with the thermal mean occupation numbers
\begin{equation}
\label{thermnumb}
n_\alpha(\nu)=\frac{1}{{\rm e}^{\beta_\alpha\nu}-1}\ .
\end{equation}

Finally, the spins will start in a
generic initial state $\rho(0)$, so that $\rho_{\rm tot}(0)=\rho(0)\otimes \rho_\beta$.

Being interested in studying the dynamics of the spin-system,
one conveniently integrates over the unobserved bath degrees of freedom 
and concentrates on the analysis of the reduced time evolution,
formally given by the transformation map:
$\rho(0)\mapsto\rho(t)\equiv{\rm Tr}_B[\rho_{\rm tot}(t)]$, where the partial trace ${\rm Tr}_B$ is computed over the bath degrees of freedom.
In the present situation, correlations in the
baths can be assumed to decay much faster than the spin-system characteristic evolution time
given by the inverse of its dominant energy scale;
a physically consistent master equation
for the reduced density matrix $\rho(t)$ can then be obtained
in the limit of vanishingly small coupling constant $\lambda$.

In practice, the dynamics of the reduced system is obtained
by suitably rescaling the time variable, $t\to t/\lambda^2$
and then taking the limit $\lambda\to 0$, following the 
mathematically precise procedure known as   {\it weak coupling limit}
\cite{Alicki-Lendi}-\cite{Merkli}.
The reduced density matrix $\rho(t)$ is then found to obey 
the following evolution equation:
\begin{equation}
{\partial\rho(t)\over \partial t}= 
{\cal H}_S[\rho]
 + \overline{{\cal D}}[\rho(t)]\ ,\quad 
{\cal H}_S[\rho]
 \equiv-i \big[H_S,\, \rho\big]\ ,
\label{reduced-equation}
\end{equation}
where 
\begin{equation}
\overline{{\cal D}}[\,\rho\,]=-\lim_{T\to\infty}{1\over T}\int_0^T d\tau\
{\cal U}_S(-\tau)\ {\cal D}'\ {\cal U}_S(\tau)\,[\, \rho\,]\ ,
\label{ergodic-mean}
\end{equation}
with unitary time-evolution given by
\begin{equation}
\label{HamDyn}
{\cal U}_S(\tau)[\rho]={\rm e}^{-i\tau H_S}\, \rho\,{\rm e}^{i\tau H_S}\ ,
\end{equation}
and second order perturbative approximation
\begin{equation}
{\cal D}'[\rho]=\lambda^2\int_0^\infty dt\ {\rm Tr}_B\Big( \big[{\rm e}^{iH_0 t}\, H'\, {\rm e}^{-iH_0 t},\big[H',\, 
\rho\otimes\rho_\beta\big]\big]\Big)\ ,
\label{dissipative-part}
\end{equation}
with $H_0$ as defined in~\eqref{totHam}. 
For the case at hand, the integrals in (\ref{ergodic-mean}) and (\ref{dissipative-part}) can be explicitly
computed and the master equation for $\rho(t)$ cast in closed form:
\begin{align}
\label{master-equation}
{\partial\rho(t)\over \partial t}=\,& {\cal H}[\rho(t)]
 + {\cal D}[\rho(t)]\equiv {\cal L}[\rho(t)]\ ,\\
&{\cal H}[\rho(t)]\equiv-i \big[H_{\rm eff},\,\rho(t)]\ .
\label{h-eff}
\end{align}
The Hamiltonian term $\cal H=\cal H_S+\cal H_{LS}$ consists of two pieces: the system Hamiltonian generator $\cal H_S$ corrected by another Hamiltonian 
generator ${\cal H}_{LS}[\rho]=-i[H_{LS}\,,\,\rho]$. Then,  an effective Hamiltonian $H_{\rm eff}=H_S + H_{LS}$ emerges that contains
a bath induced Lamb-shift contribution $H_{LS}$ besides the starting system Hamiltonian.
On the other hand, the dissipative part $\cal D$ takes a standard Gorini-Kossakowski-Sudarshan-Lindblad form, whence
the dynamical semigroup generated by~\eqref{master-equation} is composed by
completely positive maps.
Instead, let us remark that direct use of the standard second order perturbative approximation ${\cal D}'[\rho]$,
so popular in applications,
often leads to physical inconsistencies resulting in a dynamics for $\rho(t)$ that in general does 
not preserve the positivity of probabilities \cite{Dumcke}.
As we shall now discuss, the explicit expressions of $H_{LS}$ and $\cal D$ depend on whether the {\it global}
or {\it local} approach is adopted in the derivation, namely, on whether in the Hamiltonian $H_S$ in~\eqref{HamDyn} one considers or not the inter-spin $XX$ interaction 
terms.

\section{Global approach}
\label{sec:global}
In deriving the master equation (\ref{master-equation}) in the {\it global} approach, 
no additional approximations are made besides those relative to the weak coupling limit. Therefore,
in order to compute the ergodic average in (\ref{ergodic-mean}) one needs to explicitly find the
spectrum and relative eigenvectors of the spin Hamiltonian $H_S$ in (\ref{spin-hamiltonian}).
The eight energy eigenvalues $E_k$ and corresponding eigenvectors $| E_k \rangle$
are collected in Appendix~\ref{appendix-a}. The spin operators $\sigma_-^{(\alpha)}$  can then be
decomposed as
\begin{equation}
\label{A-op}
A_\alpha(\omega)=\sum_{E_\ell-E_k=\omega} \vert E_k\rangle\langle E_k\vert\,\sigma_-^{(\alpha)}\,\vert E_\ell\rangle\langle E_\ell|\ ,
\end{equation}
where the sum $\Sigma$ is over all energies eigenvalues $E_k$ and $E_\ell$ with a fixed energy difference $\omega$.
Under the working assumption that $\Delta > \sqrt{2} g$ which avoids degeneracies, the allowed values of $\omega$ are the following positive ones
\begin{equation}
\label{omegas}
\omega_0= 2\Delta\ ,\ \omega_1 = 2(\Delta +\sqrt{2} g)\ ,\ \omega_2 = 2(\Delta -\sqrt{2} g)\ ,
\end{equation}
and their negative counterparts $-\omega_i$, $i=0,1,2$. Altogether, they are such that
$\sum_\omega A_\alpha(\omega) = \sigma_-^{(\alpha)}$ as implied by $\sum_{k=1}^8\vert E_k\rangle\langle E_k\vert=1$, while 
$[ H_S, A_\alpha(\omega)]=-\omega\, A_\alpha(\omega)$.
Using the operators $A_\alpha(\omega)$, the interaction Hamiltonian in (\ref{interaction-h}) reads 
\begin{equation}
H'=\sum_{\alpha=L,R} \sum_{\omega} \Big( A_\alpha(\omega) B^\dag_\alpha + A_\alpha^\dag(\omega) B_\alpha \Big)\ .
\label{interaction}
\end{equation}
Inserting $H'$ into~\eqref{dissipative-part} and the latter expression into~\eqref{ergodic-mean}, environment correlation functions appear; due to the form~\eqref{Gibbs} 
of the environment state $\rho_\beta$, the only non vanishing correlations are the following ones 
\begin{align}
\nonumber
G_\alpha(\pm t)&\equiv{\rm Tr}_B\Big(\rho_\beta\,B_\alpha(\pm t) B_\alpha^\dag\Big)\\
\label{Ga}
&=\int_0^{+\infty}{\rm d}\nu\,{\rm e}^{\mp it\nu}\,[h_\alpha(\nu)]^2\,(1+n_\alpha(\nu))\ ,\\
\nonumber
\widetilde G_\alpha(\pm t)&\equiv{\rm Tr}_B\Big(\rho_\beta\,B_\alpha^\dag(\pm t) B_\alpha\Big)\\
\label{Gta}
&=\int_0^{+\infty}{\rm d}\nu\,{\rm e}^{\pm it\nu}\,[h_\alpha(\nu)]^2\,n_\alpha(\nu)\ .
\end{align}
where $B_\alpha(\pm t)={\rm e}^{\pm itH_B}\,B_\alpha\,{\rm e}^{\mp itH_B}$.
One then sees that, because of the ergodic average in~\eqref{ergodic-mean}, the environment influences the reduced dynamics
of the spin chain via the ``half Fourier'' transforms
\begin{equation}
\label{HFT}
\int_0^{+\infty}{\rm d}t\,{\rm e}^{\pm it\omega}\,G_\alpha(\pm t)\ ,\  \int_0^{+\infty}{\rm d}t\,{\rm e}^{\pm it\omega} \widetilde{G}_\alpha(\mp t)\ .
\end{equation}
Then, using that, in a distributional sense,
\begin{equation}
\label{distr}
\int_0^{+\infty}{\rm d}t\, {\rm e}^{\pm it(\omega-\nu)}=\mp\,i\,P\,\frac{1}{\omega-\nu}+\pi\,\delta(\nu-\omega)\ ,
\end{equation}
where $P$ denotes the principal value, the dissipative term in the master equation (\ref{master-equation}) is collected from the action of the Dirac deltas when inserted in~\eqref{HFT}. It reads:
\begin{equation}
{\cal D}[\rho]=\lambda^2 \sum_{\alpha=L,R} \ \sum_{\omega=\omega_{0,1,2}}\ {\cal D}^{(\alpha)}_{\omega}[\rho]\ ,
\label{global-dissipation-1}
\end{equation}
with
\begin{align}
\nonumber
&{\cal D}^{(\alpha)}_\omega[\rho]= C^{(\alpha)}_\omega \bigg[ A_\alpha(\omega) \rho A_\alpha^\dag(\omega)
-\frac{1}{2}\bigg\{A_\alpha^\dag(\omega)A_\alpha(\omega), \rho\bigg\}\bigg]\\
&\hskip .1cm
 +\widetilde C^{(\alpha)}_\omega \bigg[ A_\alpha^\dag(\omega) \rho A_\alpha(\omega)
-\frac{1}{2}\bigg\{A_\alpha(\omega)A_\alpha^\dag(\omega), \rho\bigg\}\bigg]\ ,
\label{global-dissipation-2}
\end{align}
where only the three positive values of $\omega$ in~\eqref{omegas} contribute because $\nu\geq 0$ in $\delta(\nu -\omega)$; explicitly,
\begin{align}
\label{kassakowski-1}
C^{(\alpha)}_\omega&= 2\pi\, [h_\alpha(\omega)]^2\,  \big(n_\alpha(\omega)+1\big)\ ,\quad \omega>0\\
\label{kassakowski-2}
\widetilde C^{(\alpha)}_\omega&=2\pi\, [h_\alpha(\omega)]^2\,  n_\alpha(\omega)\ ,\quad \omega>0\ .
\end{align}
On the other hand,  from the action of the principal value in~\eqref{distr} when inserted in~\eqref{HFT}, one gets  
the Lamb-shift correction ${\cal H}_{LS}$ to the Hamiltonian contribution $\cal H$ in~\eqref{h-eff}. It amounts to ($-i$) the commutator with the following 
Hamiltonian:
\begin{equation}
H_{LS} =  \lambda^2 \sum_{\alpha=L,R} \sum_{\omega} \left(S^{(\alpha)}_{\omega} A_\alpha^\dag(\omega)A_\alpha(\omega)
+ \widetilde S^{(\alpha)}_{\omega} A_\alpha(\omega)A_\alpha^\dag(\omega)\right)\ ,
\label{global-lamb}
\end{equation}
where the sum runs over all  positive and negative $\omega$'s and the coefficients $S^{(\alpha)}(\omega)$ and $\widetilde S^{(\alpha)}(\omega)$ read
\begin{align}
\label{Lamb1}
S^{(\alpha)}_\omega&= P\,\int_{0}^{+\infty} {\rm d}\nu\,[h_\alpha(\nu)]^2 \frac{1+n_\alpha(\nu)}{\omega-\nu}\ ,\\
\label{Lamb2}
\widetilde S^{(\alpha)}_\omega&= P\,\int_{0}^{+\infty} {\rm d}\nu\,[h_\alpha(\nu)]^2 \frac{n_\alpha(\nu)}{\nu-\omega}\ .
\end{align}
Notice that the operators $A_\alpha(\omega)$ and $A^\dag_\alpha(\omega)$ are dimensionless, whence the coefficients $C^{(\alpha)}_\omega$,  
$\widetilde{C}^{(\alpha)}_\omega$and 
$S_\omega^{(\alpha)}$ and $\widetilde{S}^{(\alpha)}_\omega$ have dimension of energy, as they should.
Furthermore, using the eigenprojections of $H_S$ and the structure of the operators 
$A_\alpha(\omega)$, one retrieves a diagonal expression for the Lamb-shift Hamiltonian:
\begin{equation}
\label{erg-exp-ham}
H_{LS}=\sum_{k=1}^8 \eta_k\,\vert E_k\rangle\langle E_k\vert\ ,\quad \eta_k\in\mathbb{R},
\end{equation}
which thus commutes with the system Hamiltonian $H_S$.

Finally, the operators $A_\alpha(\omega)$ appearing
in (\ref{global-dissipation-2}) and (\ref{global-lamb}), the so-called Lindblad operators, are explicitly given by:
\begin{align}
  &{A}_{\text{L}}(\omega_{0})=\frac{1}{2}\left(\sigma_{-}^{(1)}-\sigma_{z}^{(1)}\sigma_{z}^{(2)}\sigma_{-}^{(3)}\right)\ ,  \cr
  &{A}_{\text{L}}(\omega_{1})=\frac{1}{4}\left(\sigma_{-}^{(1)}-\sqrt{2}\sigma_{z}^{(1)}\sigma_{-}^{(2)}+\sigma_{z}^{(1)}\sigma_{z}^{(2)}\sigma_{-}^{(3)}\right)\ , \cr
  &{A}_{\text{L}}(\omega_{2})=\frac{1}{4}\left(\sigma_{-}^{(1)}+\sqrt{2}\sigma_{z}^{(1)}\sigma_{-}^{(2)}+\sigma_{z}^{(1)}\sigma_{z}^{(2)}\sigma_{-}^{(3)}\right)\ , \cr
  \label{lindblad-operators}
  &{A}_{\text{R}}(\omega_{0})=\frac{1}{2}\left(\sigma_{-}^{(3)}-\sigma_{-}^{(1)}\sigma_{z}^{(2)}\sigma_{z}^{(3)}\right)\ ,  \cr
  &{A}_{\text{R}}(\omega_{1})=\frac{1}{4}\left(\sigma_{-}^{(3)}-\sqrt{2}\sigma_{-}^{(2)}\sigma_{z}^{(3)}+\sigma_{-}^{(1)}\sigma_{z}^{(2)}\sigma_{z}^{(3)}\right)\ , \cr
  &{A}_{\text{R}}(\omega_{2})=\frac{1}{4}\left(\sigma_{-}^{(3)}+\sqrt{2}\sigma_{-}^{(2)}\sigma_{z}^{(3)}+\sigma_{-}^{(1)}\sigma_{z}^{(2)}\sigma_{z}^{(3)}\right)\ , \cr
\end{align}
for $\omega_i>0$, while the expressions for negative $-\omega_i$ are obtained form $A_\alpha(-\omega_i)=A^\dag_\alpha(\omega_i)$, $i=0,1,2$.

%
%
%
%
%
Notice that the operators $A_\alpha(\omega)$ are non local, as they couple different spin sites:
as we will see, they induce bath driven excitation transfer among different sites.

\subsection{Spin transport properties}
\label{stp:sec}
To study the transport properties of the system, we shall concentrate on the
rate of change in time of the average of $\sigma^{(2)}_z$, that is on the quantity 
\begin{equation}
\frac{d}{dt}\text{Tr}\big[ \sigma_z^{(2)}\rho(t)\big]=
\frac{d}{dt}\text{Tr}\big[ \sigma_z^{(2)}(t)\rho(0)\big]\ ,
\label{spin-flux}
\end{equation}
where in the second equality the time-evolution has been conveniently
transferred to the spin operator. In fact, the system dynamics can be equivalently formulated
in terms of evolving spin observables ${\cal O}(t)$ for any fixed initial state $\rho(0)$;
the spin observables obey the so-called ``dual'' master equation, obtained from (\ref{master-equation})
through the identity  
$\langle {\cal O}\rangle\equiv\text{Tr}\big[ {\cal O}\rho(t)\big]=
\text{Tr}\big[ {\cal O}(t)\rho(0)\big]$, valid for any initial state $\rho(0)$, so that, in general:
\begin{equation}
{\partial{\cal O}(t)\over \partial t}= i \big[H_{\rm eff},\, {\cal O}\big]
 + \widetilde{\cal D}[{\cal O}]\equiv \widetilde{\cal L}[[{\cal O}(t)]]\ ,
\label{dual-master-equation}
\end{equation}
with
\begin{equation}
\widetilde{\cal D}[{\cal O}]=\lambda^2 \sum_{\alpha=L,R} \sum_{i=0}^2\ \widetilde{\cal D}^{(\alpha)}_{\omega_i}[{\cal O}]\ ,
\label{dual-global-dissipation}
\end{equation}
\begin{align}
\nonumber
&\widetilde{\cal D}^{(\alpha)}_\omega[{\cal O}]= C^{(\alpha)}_\omega \bigg[ A_\alpha^\dag(\omega) {\cal O} A_\alpha(\omega)
-\frac{1}{2}\bigg\{A^\dag_\alpha(\omega) A_\alpha(\omega), {\cal O}\bigg\}\bigg]\\
&\hskip -.1cm 
+\widetilde C^{(\alpha)}_\omega \bigg[ A_\alpha(\omega) {\cal O} A_\alpha^\dag(\omega)
-\frac{1}{2}\bigg\{A_\alpha(\omega)A_\alpha^\dag(\omega), {\cal O}\bigg\}\bigg].
\end{align}

The Hamiltonian contribution to the rate of change in time of the average of $\sigma_z$, namely the one obtained from the first piece in the r.h.s. of 
(\ref{dual-master-equation}), can be expressed in terms of the following dimensionless operator spin currents:
\begin{align}
J^{(\ell,\ell+1)}
=4i\Big(\sigma_-^{(\ell)}\sigma_+^{(\ell+1)}-\sigma_+^{(\ell)}\sigma_-^{(\ell+1)}\Big)\ ,\ \ \ell=1,2\ ,\ \
\label{spin-current}
\end{align}
as
\begin{equation}
i\Big[H_{\rm eff},\sigma_z^{(2)}\Big]= (g+\kappa)\,\Big(J^{(1,2)}-J^{(2,3)}\Big)\ ,
\label{current-difference}
\end{equation}
where the Lamb shift contribution is characterized by a constant
\begin{equation}
\label{current-difference2}
\kappa = \frac{i\lambda^2}{8\sqrt{2}}\sum_{\alpha=L,R}\sum_{\ \ \omega=\pm\omega_1,\pm\omega_2}\Big(S_\omega^{(\alpha)}-
\widetilde S_\omega^{(\alpha)}\Big)\ .
\end{equation}
Notice that the operator differences in~\eqref{current-difference} contribute to the continuity equation as 
current divergence terms with the right dimension of energy.

Furthermore, it turns out that $\omega_0$ is not contributing to the $\omega$ sum.
An analogous behaviour holds for the dissipative contribution, as 
$\widetilde{\cal D}^{(L,R)}_{\omega_0}\big[\sigma_z^{(2)}\big]\equiv\,0$,
while for the remaining two values one has (the plus sign refers to $\omega_1$, the minus sign to $\omega_2$):
\begin{align}
\nonumber
&\widetilde{\cal D}^{(\alpha)}_{\omega}\big[\sigma_z^{(2)}\big]=
-\frac{\pi\, [h_\alpha(\omega)]^2}{2}\bigg\{\bold{1}+\Big(1+2\,  n_\alpha(\omega)\Big)\\
&\hskip 1cm\times\bigg[\sigma_z^{(2)}
\pm\frac{1}{\sqrt{2}}\Big(Q^{(1,2)}+Q^{(2,3)}\Big)\bigg]\bigg\}\ ,
\end{align}
with dimensionless operators
\begin{equation}
Q^{(i,i+1)}=\sigma^{(i)}_-\sigma^{(i+1)}_+ + \sigma^{(i)}_+\sigma^{(i+1)}_-\ ,\quad i=1,2\ .
\end{equation}
The rate of change in time of the average of $\sigma^{(2)}$ in~(\ref{spin-flux}) gives finally rise to the following continuity equation:
\begin{align}
\frac{d}{dt}\text{Tr}\big[ \sigma_z^{(2)}\rho(t)\big]=(g+\kappa)&\,\text{Tr}\Big[\big(J^{(1,2)}-J^{(2,3)}\big)\rho(t)\Big]\\
&+\text{Tr}\Big[\big(\mathcal{Q}_L+\mathcal{Q}_R\big)\rho(t)\Big]\ ,
\label{flux}
\end{align}
with operators of dimension of energy
\begin{equation}
\mathcal{Q}_\alpha=\lambda^2 \sum_{\omega=\omega_1,\omega_2}\ \widetilde{\cal D}^{(\alpha)}_\omega\big[\sigma_z^{(2)}\big]\ ,
\quad \alpha=L,R\ .
\end{equation}
One thus sees that, besides the current divergence contributions, the continuity equation~\eqref{flux} contains also extra terms that are due to the presence 
of the two heat baths; these terms cannot be cast as current differences and are interpretable as source, respectively sink contributions, depending on whether they are positive or negative. 
Furthermore, their non-vanishing is due to the global features of the Lindblad operators in~\eqref{lindblad-operators} that involve all spins of the chain: were the $A_\alpha(\omega)$  depending only on the leftmost and rightmost spin operators, the sink and source terms would disappear so that they mark a striking physical difference with respect to the local approach to be discussed in Section~\ref{sec:local}. Notice that this argument explains why in the case of just two spins as in~\cite{Levy} no sink and source contributions appear.

\subsection{Steady state}
\label{sec:global-b}
Although the master equation (\ref{master-equation}), or equivalently (\ref{dual-master-equation}), does not
allow for a simple analytic solution, it admits a unique steady state, that we will explicitly compute,
so that the asymptotic expression of the rate of change in time of the average of 
$\sigma^{(2)}_z$ in~\eqref{flux} can be accessed analytically and studied numerically.

The uniqueness of the steady state can be easily established by recalling that this is the case for all master equations
for which the commutant of ({\it i.e.} the operators commuting with)
the set of the corresponding Lindblad operators turns out to be the identity \cite{Spohn2}-\cite{Fagnola2}.
In the present case, it is convenient to work in the system energy eigenbasis.
A generic system operator can then be written
as $X=\sum_{k,\ell=1}^8 x_{k\ell}\, |E_k\rangle\langle E_\ell|$, 
so that the Lindblad operators listed in (\ref{lindblad-operators})
in the spin `computational basis' can be re-expressed in the energy eigenbasis  
as reported in Appendix~\ref{appendix-b}.
By explicit computation, one then shows that the only matrix $X$ commuting with all the elements in (\ref{lindblad-operators}) is a multiple
of the identity since the entries $x_{k\ell}$ become then of the form $x_{k\ell}=\lambda\delta_{k\ell}$, with  a same $\lambda\in\mathbb{C}$.
To obtain the explicit form of the steady state $\rho_\infty$, one has to impose the vanishing of the
r.h.s. of the master equation (\ref{master-equation}),
\begin{equation}
\label{steadystate}
{\cal L}[\rho_\infty]\equiv {\cal H}[\rho_\infty] + {\cal D}[\rho_\infty]=\,0\ .
\end{equation}
Using the expressions of the operators $A_\alpha(\omega)$ in terms of the matrix units $\vert E_j\rangle\langle E_k\vert$ constructed by means of the eigenvectors of $H_S$ as 
given in~\eqref{Eq:LingGlobEn} of Appendix~\ref{appendix-b}, one finds that $\cal D$ maps the $H_S$ eigenprojections 
into linear combinations of themselves.
Therefore, asking that ${\cal D}[\rho_\infty]=0$ on  
\begin{equation}
\label{rog}
\rho_\infty=\sum_{k=1}^8 \mu_k | E_k\rangle\langle E_k |\ ,\quad \mu_k\in\mathbb{R}\ ,
\end{equation}
namely on a matrix diagonal with respect to the $H_S$ eigenbasis, amounts to solving a system consisting of $8$ linear equations in the real unknowns $\mu_k$.
As shown in Appendix~\ref{appendix-b}, the coefficients $\mu_k$ can be grouped in the following vector:
\begin{equation}
\label{muvec}
\vec\mu=\frac{1}{(s_0+\tau_0)(s_1+\tau_1)(s_2+\tau_2)}\begin{pmatrix}
    \tau_0\tau_1\tau_2 \cr
    s_0s_1s_2 \cr
    s_0\tau_1\tau_2 \cr 
    \tau_0s_1s_2 \cr 
    \tau_0\tau_1s_2 \cr
    s_0s_1\tau_2 \cr 
    s_0\tau_1s_2 \cr 
    \tau_0s_1\tau_2
\end{pmatrix}\ ,
\end{equation}
where the steady state eigenvalues $\mu_k$ involve the quantities $\tau_i:=\tau(\omega_i)$ and $s_i:=s(\omega_i)$ with
\begin{align}
\label{tom}
&\tau(\omega)=\sum_{\alpha=L,R}\big[h_\alpha(\omega)\big]^2 n_\alpha(\omega)\ ,\\
\label{som}&s(\omega)=\sum_{\alpha=L,R}\big[h_\alpha(\omega)\big]^2 \big(n_\alpha(\omega)+1\big)\ ,
\end{align}
where $n_\alpha(\omega)$ are the mean thermal occupation numbers  in~\eqref{thermnumb}.
Furthermore, due to the diagonal form~\eqref{erg-exp-ham} of the Lamb-shift Hamiltonian, it also turns out that ${\cal H}[\rho_\infty]=0$. Then, if the coefficients $\mu_k$ determined by ${\cal D}[\rho_\infty]=0$ are positive, by the uniqueness of the stationary state, $\rho_\infty$ as in~\eqref{rog} solves~\eqref{steadystate}
(see Appendix~\ref{appendix-b}  for details).
A special case worth mentioning here is when the two baths are identical. Then, $T_L=T_R=T$  and $h_{L,R}(\omega)=h(\omega)$,
so that
\begin{equation}
\label{Gibbs1}
\hskip-.2cm
\tau(\omega)=2\,\big[h(\omega)\big]^2\, n(\omega)\ ,\
s(\omega)=2\,\big[h(\omega)\big]^2 \big(n(\omega)+1\big)\ , 
\end{equation}
where we set $n(\omega):=n_L(\omega)=n_R(\omega)$.
In such a case, the smearing function $h(\omega)$ disappears from the components of the vector~\eqref{muvec} and the partition function (normalization factor)
reduces to
\begin{equation}
\label{Gibbs3}
Z_\beta={\rm e}^{-3\,\beta\,\Delta}\,\sum_{k=1}^8{\rm e}^{-\beta\,E_k}\ .
\end{equation}
It thus follows that, for identical baths, the stationary state is the Gibbs equilibrium state at the baths temperature: indeed, one explicitly computes
\begin{equation}
\label{Gibbs4}
\mu_k=\frac{{\rm e}^{-\beta\,E_k}}{\sum_{\ell=1}^8{\rm e}^{-\beta\, E_\ell}}\quad\hbox{whence}\quad \rho_\infty=\frac{{\rm e}^{ -\beta\,H_S}}{Z_\beta}\ .
\end{equation}
In general, that is when the baths differ either in temperature or in the smearing functions $h_{L,R}(\omega)$, the stationary state is no longer thermal with respect to $H_S$.

\begin{figure}
    \centering
    \includegraphics[width=1\linewidth]{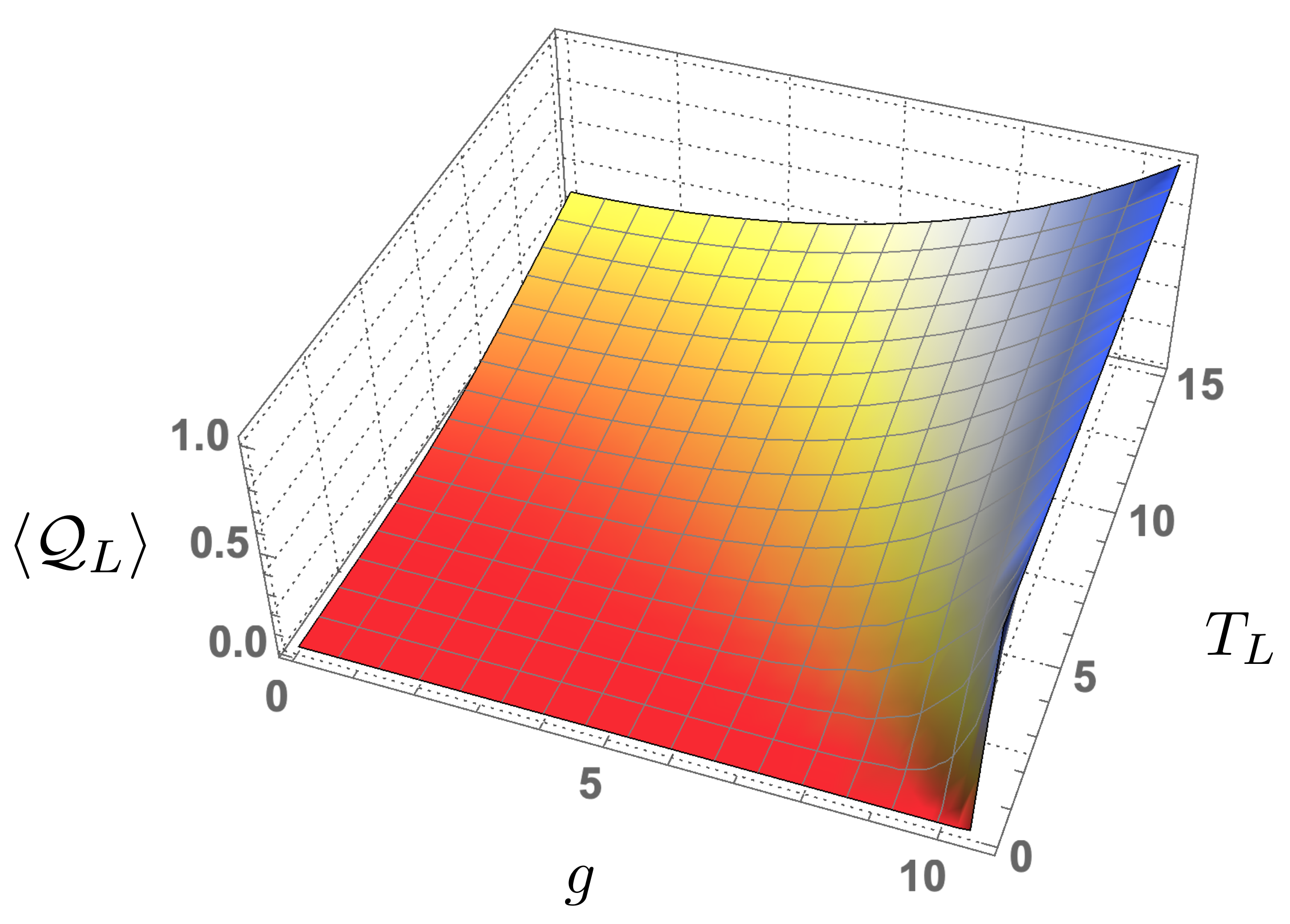}
    \caption{Steady state average source contribution, $\langle\mathcal{Q}_L \rangle$ (in dimension of energy),
as a function of the left bath temperature $T_L$ and the inter-spin coupling constant $g$,
with the right bath temperature set to $T_R=\,0$ and $\Delta=15$, in units of $\pi\lambda^2h^2/4$.
For small values of $g$, the source contribution remains small even
for large bath temperature differences; on the contrary, as soon as 
$g$ becomes comparable with $\Delta$ the source/sink terms cannot be ignored.
(See also subsequent figure.)
}
    \label{Fig2a}
\end{figure}

\begin{figure}
    \centering
    \includegraphics[width=1\linewidth]{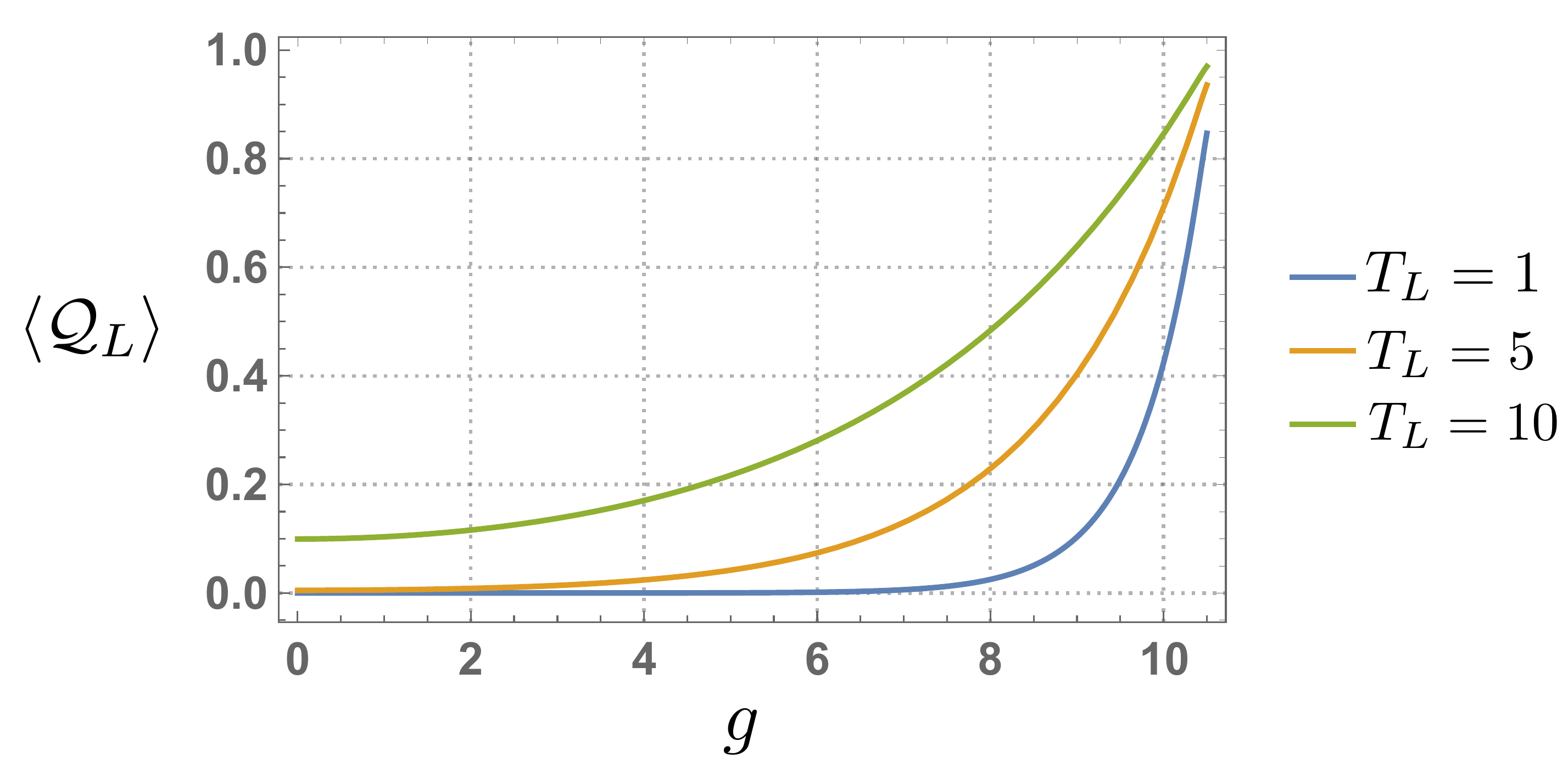}
    \caption{Steady state average source contribution, $\langle\mathcal{Q}_L \rangle$ (in dimension of energy), 
as a function of the inter-spin coupling constant $g$ for fixed values
of the left bath temperature $T_L$; these are slices of the previous figure
along the indicated three $T_L={\rm constant}$ planes.}
    \label{Fig2b}
\end{figure}

With the explicit steady state at disposal, one can now study the fate of the various contributions
to the rate of change in time of the average of $\sigma^{(2)}_z$
in~\eqref{flux} for asymptotically long times.
First of all, as the spin-currents $J^{(i,i+1)}$ in (\ref{spin-current}) have zero
expectations with respect to the energy eigenstates $|E_k\rangle$, they vanish in the steady state:
\begin{equation}
{\rm Tr}\Big[J^{(1,2)}\,\rho_\infty\Big]=\Big[J^{(2,3)}\,\rho_\infty\Big]=\,0\ .
\end{equation}
Instead, for the sink/source terms, using~\eqref{tom} and~\eqref{som}, one finds:
\begin{align}
\nonumber
&\text{Tr}\big[\mathcal{Q}_{L}\, \rho_\infty\big]=\frac{\lambda^2\,\pi}{4} 
\sum_{\omega=\omega_1,\omega_2}\big[h_L(\omega)\big]^2\\
&\hskip 1cm
\times\frac{n_L(\omega)\big[s(\omega)-\tau(\omega)\big]-\tau(\omega)}{s(\omega)+\tau(\omega)}\ ,
\label{source}
\end{align}
while the $R$-contribution is exactly the opposite,
\begin{equation}
\text{Tr}\big[\mathcal{Q}_{R}\, \rho_\infty\big]=-\text{Tr}\big[\mathcal{Q}_{L}\, \rho_\infty\big]\ ,
\end{equation}
as it should be in a steady state where the rate of change in time of the average of $\sigma^{(2)}_z$ must vanish~\cite{Prosen5}.
In the simplified case for which the smearing functions $h_\alpha(\omega)$ introduced in (\ref{bath-operator})
are the same for both $L$ and $R$ baths as in the case of identical baths, by further setting $h_\alpha(\omega_1)=h_\alpha(\omega_2)=h$,
the result in (\ref{bath-operator}) reduces to
\begin{equation}
\label{ss1}
\text{Tr}\big[\mathcal{Q}_{L}\, \rho_\infty\big]=\frac{\lambda^2 h^2\,\pi}{4}
\sum_{\omega=\omega_1,\omega_2}\frac{{n}_L(\omega)-{n}_R(\omega)}{{n}_L(\omega)+{n}_R(\omega)+1}\ .
\end{equation}
When the $L$-bath temperature is higher than the one of the $R$-bath, then $\beta_L\leq\beta_R$ whence $n_L(\omega)\geq n_R(\omega)$
so that $\text{Tr}\big[\mathcal{Q}_{L}\, \rho_\infty\big]\geq0$ becomes a source term and $\text{Tr}\big[\mathcal{Q}_{R}\, \rho_\infty\big]\leq0$ a sink term.
Clearly, the roles are interchanged for the reverse temperature hierarchy, $T_R\leq T_L$.\\
As depicted in Fig.~\ref{Fig2a}, at fixed $g$, with increasing difference between the bath temperatures, the source contribution initially grows and then saturates to 
$\displaystyle{\frac{\lambda^2\pi h^2}{2}}$. Indeed, $T_R=0$ sets $n_R(\omega)=0$ and $\displaystyle n_L(\omega)\simeq \frac{1}{\beta_L\,\omega}$ when $\beta_L\to0$. The expression~\eqref{ss1} depends continuously on $g$ 
through the frequencies $\omega_{1,2}=2(\Delta\pm\sqrt{2}g)$ and in the limit of vanishing  $g$ one gets 
\begin{equation}
\label{ss2}
\text{Tr}\big[\mathcal{Q}_{L}\, \rho_\infty\big]=
\frac{\lambda^2 h^2\,\pi}{4}\ \frac{{\rm e}^{\beta_L\Delta}-{\rm e}^{\beta_R\Delta}}{{\rm e}^{\beta_L\Delta}+{\rm e}^{\beta_R\Delta}}\ .
\end{equation}
Therefore, though they disappear as they should when $T_L=T_R$, sink and source terms are nevertheless present even in the limit of vanishing  inter-spin interactions.
As observed at the end of Section~\ref{stp:sec}, this is physically untenable since then the global features of the Lindblad operators~\eqref{lindblad-operators} should disappear 
and one would expect only gradient-like contributions as those emerging from the local approach discussed in the next section (see~\eqref{local-flux}).
The reason for the presence of sink/source contributions at $g=0$ is that,  in the global approach, the technical machinery providing the reduced dynamics is not justified because of the degeneracy of the spin transition frequencies. Indeed, the global character of the Lindblad operators $A_\alpha(\omega)$ being independent on $g$ shows that the time-limit in the ergodic average~\eqref{ergodic-mean} leading to the master equation and the $g\to0$ limit corresponding to switching off the inter-spin interactions cannot be interchanged. There is a discontinuity of the two approaches with respect to a vanishing coupling constant; however,  as shown in Fig.~\ref{Fig2a} and Fig.~\ref{Fig2b} this effect becomes relevant only when $g\simeq \Delta$, namely away from degeneracy, and vanishes as soon as the difference between the temperatures of the two baths goes to zero.


\section{Local approach}
\label{sec:local}
In the local approach, the derivation of the master equation is simplified, as the inter-spin interaction
is assumed to be negligible with respect to the couplings to both the transverse magnetic field and the baths; in other terms, one assumes 
$g\ll \Delta, \lambda B$ where $B$ stands for the smaller of the norms of the operators $B_{L,R}$ in Eq.~\eqref{bath-operator}.
Therefore, the $L$-bath interacts with the first spin of the chain, while the $R$-bath with the third, as if the two end spins were isolated from the middle one, and
the inter-spin interaction is switched back on only after the weak-coupling procedure has been applied. Within this approach, one thus performs the ergodic average in~\eqref{reduced-equation} using a unitary evolution in~\eqref{HamDyn} 
with the Hamiltonian $H_S$ in~\eqref{spin-hamiltonian} replaced by $\displaystyle H_S=\Delta\,\sum_{i=1}^3\sigma^{(i)}_z$.
The resulting master equation is again
of the form (\ref{master-equation}), with a dissipative term $\cal D$ which
is the sum of two similar bath contributions $(\alpha=L,R$),
\begin{align}
&\mathcal{D}_\alpha[\rho]=2\lambda^2\,\pi\, [h(\Delta)]^2\, n_\alpha\bigg[\sigma_+^{(\alpha)}\rho\,\sigma_-^{(\alpha)}
-\frac{1}{2}\Big\{\sigma_-^{(\alpha)}\sigma_+^{(\alpha)},\rho\Big\}\bigg]\cr
&\hskip-.1cm+\,2\,\lambda^2\,\pi \,[h(\Delta)]^2\,(1+n_\alpha)\bigg[\sigma_-^{(\alpha)}\rho\,\sigma_+^{(\alpha)}-\frac{1}{2}\Big\{\sigma_+^{(\alpha)}\sigma_-^{(\alpha)},\rho\Big\}\bigg]\ ,
\label{local-dissipation}
\end{align}
where the definitions in (\ref{sigmas}) have been used while we have set  $n_\alpha:=n_\alpha(\Delta)$ (see~\eqref{thermnumb}) for the only two 
contributing thermal occupation numbers.
Further, for simplicity, we have chosen $h_L=h_R\equiv h$ in the factors coming from the Fourier transforms of the thermal correlation functions 
(compare with (\ref{kassakowski-1}) and~\eqref{kassakowski-2}).
Similarly, the Lamb-shift contributions
to the Hamiltonian piece can be reabsorbed in a redefinition of the constant magnetic field
strength $\Delta$, so that in practice $H_{\rm eff}=H_S$, but now with $g\neq0$.
As a result, the local approach yields  the following 
master equation for the spin density matrix:
\begin{equation}
{\partial\rho(t)\over \partial t}=-i \big[H_S,\,\rho(t)]
 + {\cal D}_L[\rho(t)] + {\cal D}_R[\rho(t)]\equiv{\cal L}[\rho(t)]\ .
\label{local-master-equation}
\end{equation}

\subsection{Spin transport properties}
In the local approach,  the transport properties of the spin chain are also addressed by looking  at the rate of change in time of the average of $\sigma_z^{(2)}$ 
by means of the definition given in (\ref{spin-flux}). 
Recalling (\ref{current-difference}), one shows that
the Hamiltonian contribution to~\eqref{spin-flux} can be
recast again in terms of the difference of the two spin-currents $J^{(1,2)}$ and $J^{(2,3)}$
defined in (\ref{spin-current}). However, no bath contributions can now arise,
as the dissipative pieces in (\ref{local-dissipation}) do not involve the middle spin.
Therefore, in the local approach, the continuity equation reads
\begin{equation}
\frac{d}{dt}\text{Tr}\big[ \sigma_z^{(2)}\rho(t)\big]=\, g\,\text{Tr}\Big[\big(J^{(1,2)}-J^{(2,3)}\big)\rho(t)\Big]\ ,
\label{local-flux}
\end{equation}
with no bath-induced sink/source terms.

\subsection{Steady state}
Although the steady states of boundary-driven $XX$ spin-chains have been studied before
in terms of {\it matrix product states}$\,$%
\footnote{See \cite{Prosen4} and references therein.},
we shall give here a more explicit description
for the specific situation at hand based on a perturbative expansion.

First of all, also the master equation (\ref{local-master-equation}) generates a relaxing dynamics,
admitting a unique steady state,
to which any initial spin state tends for asymptotically long times. This result can be easily proven
using the same strategy adopted in Section \ref{sec:global-b} for the global approach dynamics.
Working again in the spin energy eigenbasis, any spin operator 
$X=\sum_{k,\ell=1}^8 x_{k\ell}\, |E_k\rangle\langle E_\ell|$ that commutes
with $H_S$ has a diagonal matrix of coefficients $x_{k\ell}$. Further, demanding
commutation with all Lindblad operators appearing in (\ref{local-master-equation}), namely
$[X,\sigma_{\pm}^{(i)}]=\,0$, $i=1,3$, imposes the coefficients $x_{k\ell}$ to form a matrix proportional 
to the identity matrix which then results the only element of the commutant of the set $\{H_S, \sigma_{\pm}^{(L)},\sigma_{\pm}^{(R)}\}$ 
and, as mentioned before, this guarantees the uniqueness of the steady state.

In order to determine the explicit expression of the steady state $\rho_\infty$, one first observes by direct inspection
that the action of the operator $\cal L$ leaves invariant the linear span
generated by the following 14 operators written in the spin `computational basis':
\begin{align}
&{\cal E}_{jk\ell}=\ket{jk\ell}\bra{jk\ell},\  j, k, \ell=0,1\ ,\cr
&{\cal F}_{1}=\ket{001}\bra{100}+\ket{100}\bra{001},\quad  {\cal F}_{2}=\sigma_x^{(2)}{\cal F}_1\sigma_x^{(2)}\ ,\cr
&{\cal F}_{3}=i(\ket{001}\bra{010}-\ket{010}\bra{001}),\  {\cal F}_{4}=\sigma_x^{(1)}{\cal F}_1\sigma_x^{(1)}\ ,\cr
&{\cal F}_{5}=i(\ket{011}\bra{101}-\ket{101}\bra{011}),\  {\cal F}_{6}=\sigma_x^{(3)}{\cal F}_1\sigma_x^{(3)}\ .\cr
\end{align}
The eight operators $\cal E$ are diagonal, while the remaining six $\cal F$ are hermitian, off-diagonal.
Clearly, the steady state $\rho_\infty$ must be a normalized, linear combination of these operators, and the 
condition ${\cal L}[\rho_\infty]=\,0$, yielding a system of 14 linear equations in the unknown coefficients,
will fix it completely. However, the expression of these coefficient turns out to be rather cumbersome
and a compact, explicit version for them hard to find.
It thus proves more convenient to seek a perturbative expression for the stationary state.

As in the local approach the coupling $g$ between the spins is considered to be small, $g\ll\Delta, \lambda B$,
we treat the spin-interaction as a perturbation and  rewrite the dynamical generator $\cal L$ in (\ref{local-master-equation}) as
\begin{equation}
\mathcal{L}=\mathcal{L}_0+g\,\mathcal{L}_1\ ,
\label{L-expansion}
\end{equation}
where
\begin{align}
\label{L0}
&{\cal L}_0[\rho]=-i \Big[\Delta\sum_{i=1}^3\sigma_z^{(i)},\,\rho\Big]
 + {\cal D}_L[\rho] + {\cal D}_R[\rho]\ ,\\
&{\cal L}_1[\rho]=-i \bigg[\sum_{i=1}^2\left(\sigma_{x}^{(i)}\sigma_{x}^{(i+1)}
+\sigma_{y}^{(i)}\sigma_{y}^{(i+1)}\right),\,\rho\bigg]\ .
\end{align}
Then, expressing the steady state $\rho_\infty$ as a power series expansion:
\begin{equation}
\rho_\infty=\sum_{n=0}^\infty g^n\,\rho^{(n)}\ ,
\end{equation}
the steady state condition ${\cal L}[\rho_\infty]=\,0$ reduces to:
\begin{equation}
\mathcal{L}_0[\rho^{(0)}]+\sum_{n=1}^\infty g^n\,\Big(\mathcal{L}_0[\rho^{(n)}]\,+\,\mathcal{L}_1[\rho^{(n-1)}]\Big)=0\ ,
\end{equation}
leading to the following recursive relations that must be satisfied for all $n$:
\begin{equation}
\label{recursive-conditions}
\mathcal{L}_0[\rho^{(0)}]=\,0\ ,\qquad \mathcal{L}_0[\rho^{(n+1)}]=-\mathcal{L}_1[\rho^{(n)}]\ .
\end{equation}
Therefore, once a stationary state $\rho^{(0)}$ of $\mathcal{L}_0$ is chosen, its first order perturbation is obtained as
\begin{equation}
\label{rho1}
\rho^{(1)}=-\mathcal{L}_0^{-1}\circ\mathcal{L}_1[\rho^{(0)}]\ ,
\end{equation}
through the inversion of $\mathcal{L}_0$, and similarly for the higher order terms:
\begin{equation}
\rho^{(n+1)}=-\mathcal{L}_0^{-1}\circ\mathcal{L}_1[\rho^{(n)}]\ .
\end{equation}
Though $\mathcal{L}_0$ is in general not invertible,  $\mathcal{L}_0^{-1}$ can be defined 
on a subspace that does not contain elements of the kernel of $\mathcal{L}_0$ (see Appendix C for further details).

In the case at hand, the state $\rho^{(0)}$ such that \hbox{$\mathcal{L}_0[\rho^{(0)}]=\,0$} is of the form:
\begin{equation}
\rho^{(0)}=\rho_L\otimes\rho_r\otimes\rho_R\ ,
\label{rho0}
\end{equation}
where, using the `computational basis' of spins as in Appendix A,
\begin{equation}
    \rho_\alpha=\frac{1}{1+2\,n_\alpha}\begin{pmatrix}
    n_\alpha&0\cr
    0&1+n_\alpha
    \end{pmatrix}\ ,\quad \alpha=L,R\ ,
\end{equation}
are thermal states, while $\rho_r$ is an arbitrary diagonal density matrix:
\begin{equation}
\rho_r=\begin{pmatrix}
r & 0\\
0 & 1-r
\end{pmatrix}\ ,\quad 0\leq r\leq 1\ .
\end{equation}
The ${\cal L}_0$-stationary state is not unique; indeed, the commutant of the Lindblad operators appearing in (\ref{L0}) is not the identity,
rather the linear span generated by the two operators:
\begin{equation}
P_\pm={\bf 1}\otimes\frac{{\bf 1}\pm\sigma_z}{2}\otimes{\bf 1}\ .
\label{p-operators}
\end{equation}

The action of the perturbation ${\cal L}_1$ on $\rho^{(0)}$ can now be straightforwardly obtained;
in the tensor product basis and with the ordering used in Appendix A, one finds (only the non-vanishing entries
are explicitly shown):
\begin{equation}
\mathcal{L}_1\big[\rho^{(0)}\big]=i\,\begin{pmatrix}
\cdot&\cdot&\cdot&\cdot&\cdot&\cdot&\cdot&\cdot\\
\cdot&\cdot&a&\cdot&\cdot&\cdot&\cdot&\cdot\\
\cdot&-a&\cdot&\cdot&b&\cdot&\cdot&\cdot\\
\cdot&\cdot&\cdot&\cdot&\cdot&c&\cdot&\cdot\\
\cdot&\cdot&-b&\cdot&\cdot&\cdot&\cdot&\cdot\\
\cdot&\cdot&\cdot&-c&\cdot&\cdot&d&\cdot\\
\cdot&\cdot&\cdot&\cdot&\cdot&-d&\cdot&\cdot\\
\cdot&\cdot&\cdot&\cdot&\cdot&\cdot&\cdot&\cdot\\
\end{pmatrix}\ ,
\label{L1}
\end{equation}
with 
\begin{align}
&a=2\,r_L(r-r_R)\ ,\cr
&b=2\,r_R(r_L-r)\ ,\cr
&c=2\,(r_L-r)(1-r_R)\ ,\cr
&d=2\,(r-r_R)(1-r_L)\ ,
\end{align}
and \vskip -.5cm
\begin{equation}
r_\alpha=\frac{n_\alpha}{1+2n_\alpha}\ ,\quad\alpha=L,R\ .
\end{equation}
One can check that the just obtained $\mathcal{L}_1\big[\rho^{(0)}\big]$
is not in the kernel of ${\cal L}_0$, so that
$\rho^{(1)}$ in (\ref{rho1}) can be safely computed and found to be of the same form
as the matrix in (\ref{L1}) but with the four constants replaced by
%
\begin{align}
&a'=\,\frac{4(r_R-r)}{1+2n_R} \frac{2n_L+r_L(1+2n_R)}{3+4n_L+2n_R}\ ,\cr
&b'=\,\frac{4(r-r_L)}{1+2n_L} \frac{2n_R+r_R(1+2n_L)}{3+2n_L+4n_R}\ ,\cr
&c'=\,\frac{4(r-r_L)}{1+2n_L}\frac{3+2n_L+2n_R-r_R(1+2n_L)}{3+2n_L+4n_R}\ ,\cr
&d'=\,\frac{4(r_R-r)}{1+2n_R} \frac{3+2n_R+2n_L-r_L(1+2n_R)}{3+4n_L+2n_R}\ .
\label{rho1-parameters}
\end{align}

The corresponding expression for $\rho^{(1)}$ should give the first-order correction in the expansion of the
the steady state $\rho_\infty$; however, while we know $\rho_\infty$ to be unique,
both $\rho^{(0)}$ and $\rho^{(1)}$ still have $r$ as a free parameter.
This situation is common in perturbation theory \cite{Baumgartner1}-\cite{Baumgartner3}: 
in order to fix $r$, 
one needs to examine the next perturbative order using (\ref{recursive-conditions}), 
and apply ${\cal L}_1$ to $\rho^{(1)}$. By requiring ${\cal L}_1[\rho^{(1)}]$ not to belong
to the kernel of ${\cal L}_0$, so that the second order perturbative contribution $\rho^{(2)}$ can
be determined, fixes uniquely the parameter $r$:
\begin{equation}
\label{value-r}
r=\frac{r_R(1+2n_L)+r_L(1+2n_R)}{2(1+n_R+n_L)}\ .
\end{equation}
Notice that $0\leq r\leq 1$, since $0\leq r_{R,L}\leq 1$. 

In conclusion, the steady state of the master equation (\ref{local-master-equation}), up to the first order 
in the coupling $g$, is given by
\begin{equation}
\rho_\infty=\rho^{(0)}+g\rho^{(1)}\ ,
\label{local-steady-state}
\end{equation}
where $\rho^{(0)}$ and $\rho^{(1)}$ are as in (\ref{rho0}) and (\ref{L1}), (\ref{rho1-parameters}),
with the parameter $r$ as in (\ref{value-r}).
Notice that for identical baths, $T_L=T_R=T$ implies 
\begin{eqnarray}
\label{Gibbs5a}
n_L=n_R=:n&=&\frac{1}{{\rm e}^{2\,\beta\,\Delta}-1}\ ,\\
\label{Gibbs5b}
r_L=r_R=r&=&\frac{n}{1+2n}\ .
\end{eqnarray}
Then, all coefficients in~\eqref{rho1-parameters} vanish and $\rho^{(1)}$ as well.
Hence, at first order in $g$, the local stationary state is a Gibbs thermal state:
\begin{equation}
\label{Gibbs6}
\rho_\infty=\rho^{(0)}= \rho_n\otimes\rho_n\otimes\rho_n=\frac{{\rm e}^{-\beta \, H_S}}{{\rm Tr}{\rm e}^{-\beta\,H_S}}\ , 
\end{equation}
where $H_S$ is the spin Hamiltonian in~\eqref{spin-hamiltonian} with $g=0$ and 
\begin{equation}
\label{Gibbs7}
\rho_n:=\frac{1}{1+2n}\begin{pmatrix}n&0\cr0&1+n\end{pmatrix}\ .
\end{equation}

A first interesting conclusion that can be drawn from 
comparing the stationary states in the global and local approaches is that the local regime does not emerge from the global one by letting $g=0$. 
Indeed, in general, already the order zero expansion with respect to $g$ of the 
stationary state $\rho_\infty$ in~\eqref{rog} derived in the global approach differs from the order zero term $\rho^{(0)}$ in the local approach.
In order to appreciate this fact, consider the coefficients given in~\eqref{bmu} in Appendix~\ref{appendix-b}: they depend on $g$ through the Hamiltonian $H_S$ eigenvalues.
By setting $g=0$, from~\eqref{omegas} one finds $\omega_1=\omega_2=\omega_3=2\Delta$, whence in the expressions~\eqref{tom} and~\eqref{som} 
$n_{L,R}(\omega)=:n_{L,R}$ and $h_{L,R}(\omega)=:h$ as already seenin the case of identical baths (see~\eqref{Gibbs1}). Then, 
one finds $\tau=x\,h^2$ and
$s=h^2(2+x)$ with $x:=n_L+n_R$, so that, as before, the factor $h$ disappears again from the coefficients $\mu_k$ in~\eqref{muvec}.
Then, the stationary  state $\rho_\infty$ becomes
\begin{equation}
\label{rg0}
\rho_\infty=\rho_x\otimes\rho_x\otimes\rho_x\, ,\quad \rho_x:=\frac{1}{2(1+x)}\begin{pmatrix}x&0\cr0&2+x\end{pmatrix}\ .
\end{equation}
Therefore, the global stationary state in~\eqref{rog} computed  in the limit of vanishing $g$  
and $\rho^{(0)}$ in~\eqref{Gibbs6} 
can coincide only for equal left and right temperatures.

Furthermore, the expectation that the two regimes correspond to different physical scenarios is strikingly confirmed when one analyzes
the asymptotic behaviour of the spin currents $J^{(1,2)}$ and $J^{(2,3)}$ that enter the expression
of the rate of change in time of the average of $\sigma_z^{(2)}$ in (\ref{local-flux}). Using (\ref{local-steady-state}), 
one finds that only the first order term $\rho^{(1)}$ in the perturbative expansion contributes to the asymptotic average of the currents:
\begin{align}
\nonumber
&{\rm Tr}\Big[J^{(1,2)}\,\rho_\infty\Big]= 8\,(b' + c')\\ 
&{\rm Tr}\Big[J^{(2,3)}\,\rho_\infty\Big]=8\,(a' + d')\ .
\label{local-currents}
\end{align}
These mean values are in general nonzero and vanish only when the two bath temperatures $T_L$ and $T_R$
are equal, since in this case, as previously seen, $a'=b'=c'=d'=0$.
In general, the total rate of change in time of the average 
of $\sigma^{(2)}_z$ in (\ref{local-flux})
is always zero, as it should be in a steady state; indeed, the two expressions in (\ref{local-currents})
are equal, as the condition $a'+d'=b' + c'$ is precisely the one that fixes the parameter $r$
to assume the value in (\ref{value-r}).

\section{Discussion}

Stimulated by the ongoing debate on the different available approaches that can be adopted for analyzing the transport
properties of open quantum systems, we have studied the 
asymptotic spin-transport properties of a three-site spin-1/2 chain,
with $XX$-type interaction in the presence of a constant magnetic field, weakly coupled at the two ends to two
separate heat baths, at different temperatures. The merit of such a simplification is that it allows for an analytical 
determination of the asymptotic states and corresponding transport properties of the spin chain. 
Master equations generating the reduced spin dynamics have been derived in the {\it weak coupling} (Markovian) limit
using both the {\it global} approach, where the full $XX$ Hamiltonian is always taken into account,
and the {\it local} approach, where instead the spin-spin interactions are neglected.
Both types of master equations admit unique stationary states, whose forms have been explicitly
derived in the global regime and up to its first order perturbation expansion with respect to the inter-spin interaction 
in the local regime, 
thus allowing a complete analytic treatment of the system asymptotic transport properties.

\begin{figure}
    \centering
    \includegraphics[width=1\linewidth]{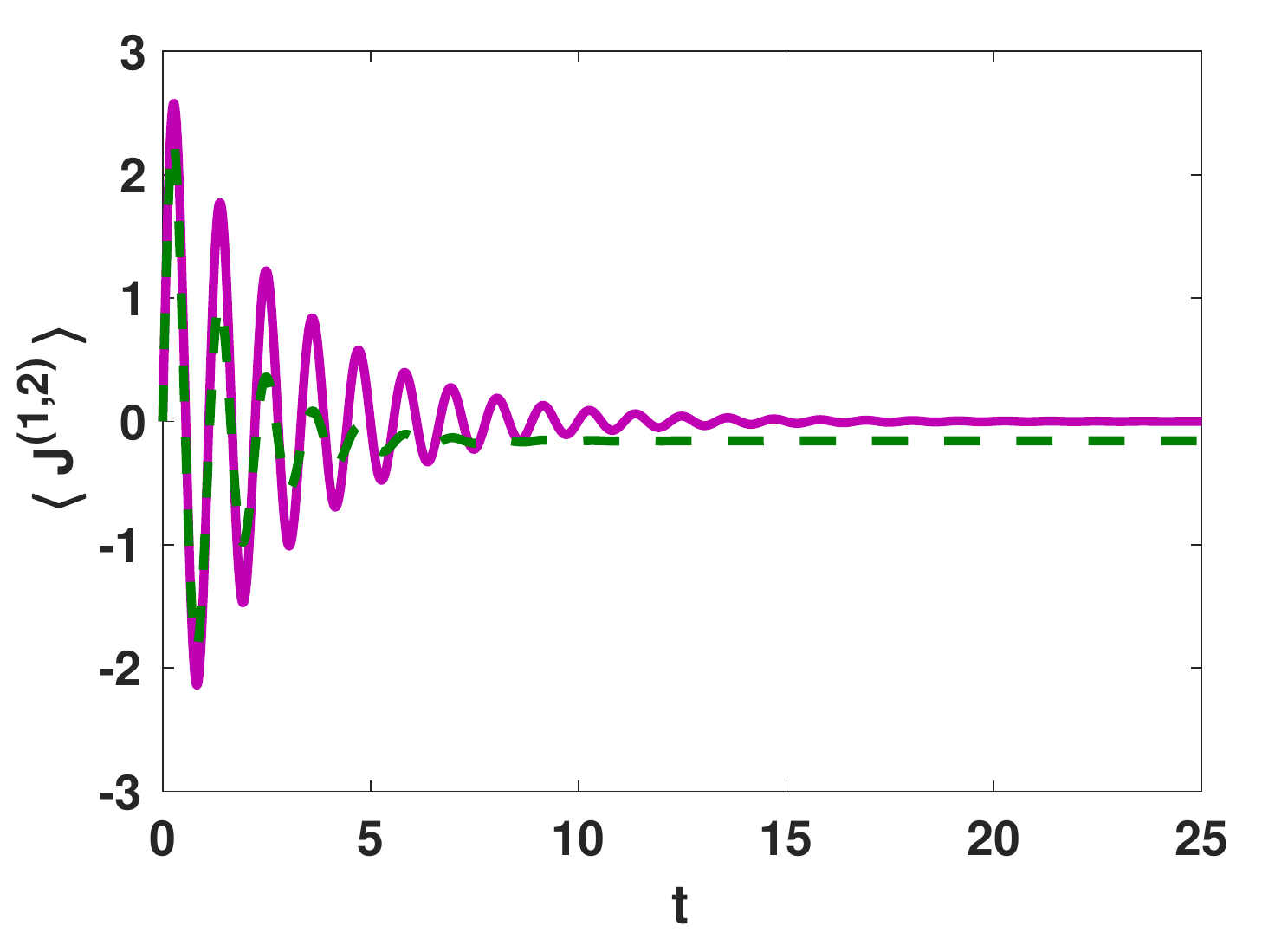}
    \caption{Average of the dimensionless spin current $J^{(1,2)}$ between sites one and two for $\lambda=1$, $g=1$, $\Delta=15$, $T_L=10$ and $T_R=20$, in natural units ($\hbar=k_{\rm Boltzmann}=1$). The solid purple line corresponds to the global approach, while the dashed green line represents the local approach; note that in the latter approach the steady state supports a non-vanishing value for the current average. A similar behaviour holds for the average of the spin current  $J^{(2,3)}$ between sites two and three.}
    \label{Fig3}
\end{figure}

In particular, we have focused on the behaviour of the rate of change in time of the average of the middle spin $z$ component $\sigma^{(2)}_z$.
Though we concentrated on the asymptotic spin-transport properties, as far as the transient dynamics is concerned,
the corresponding continuity equation allows defining spin currents involving the first two sites, $J^{(1,2)}$,
whose damped (due to the baths) oscillatory (due to the Hamiltonian) 
behaviour is depicted in Fig.~\ref{Fig3} 
in both approaches. For the last two spins the behaviour of the spin current  $J^{(2,3)}$ is similar.
Notice that in contrast with the {\it global} case, the {\it local} master equation supports 
non vanishing asymptotic values for the two spin currents, becoming zero when the bath temperatures
are equal and, obviously, when the inter-spin coupling vanishes; these values are nevertheless equal so that the asymptotic global 
rate of change in time of the average of $\sigma^{(2)}_z$ is zero,
as it should be in a stationary state. 
The different time-behaviours of the spin-currents as shown in Fig.~\ref{Fig3} could be helpful in sorting out
the validity range of the two approaches.

In all cases, the most striking difference between the local and global approach to the description of the open chain spin transport properties
comes from the presence of additional bath induced pieces in the {\it global} approach continuity equation that correspond to sink and source contributions.
Their origin can be traced to the $XX$-type self-coupling among the spins: this interaction is fully taken into account by the
{\it global} master equation, that indeed allows bath assisted 
global effects by virtue of Lindblad operators that involve all three spins, unlike in the local approach where they refer only to the leftmost and rightmost spins, namely those coupled directly to the heat baths.
However, as shown by Figs.~\ref{Fig2a} and~\ref{Fig2b}, though small, sink and source contributions remain even at vanishing inter-spin coupling, whereas one would expect them to vanish due to the local structure of the Lindblad operators: this phenomenon is due to the lack of commutativity between the time-limit in the ergodic average leading to the Lindblad master equation in its weak-coupling limit derivation and the limit in which the inter-spin coupling constant is left to vanish. Such a lack of interchangeability of the limits is clearly put into evidence  by the fact that the expansion of the steady state in the 
global regime with respect to the inter-spin coupling does not lead to the first order approximation of the 
steady state in the local regime, both of them being Gibbs thermal states with respect to the corresponding spin Hamiltonians, namely with, respectively without inter-spin interactions,  when the bath temperatures are equal.
The results presented above refer to the asymptotic transport properties in the global and local approaches, as such they cannot offer indications about the different time-scales present in the transient dynamics of the spin-chain. However, they point to a physical discontinuity between the two regimes, this means that for sufficiently small values of the inter-spin coupling the local approach is the only one valid, while, for sufficiently large ones, the global approach is the tenable one, with probably a range of values where both approaches together would contribute to a proper description (see for instance~\cite{Giovannetti20}). Indeed, both regimes are dissipative approximations, on their own proper time-scales, of the reduced spin dynamics resulting from the true reversible global dynamics of chain and environment together.  The extension of the validity regions and their possibile overlap can only be determined by a thorough investigation of the transient dynamics, numerical or experimental, a task to which the presence of the sink and source contributions  and of the currents and of their asymptotic values in the two regimes, as from Fig.~\ref{Fig3}, certainly lend concrete and interesting motivations.


Finally, we briefly compare the previous results with those obtained in~\cite{Levy} and~\cite{Trushechkin} for the heat currents  in an open chain consisting of two spins.
According to the standard approach to quantum thermodynamics~\cite{Alicki1}--~\cite{Spohn1}, in the steady state $\rho_\infty$, the heat currents flowing between the thermal baths and the spin chain are given by
\begin{equation}
\label{eq:HeatCurrent}
J_Q^{(\alpha)}(\rho_\infty)=-\lambda^2\text{Tr}\left(H_S\sum_{\omega_0,\omega_1,\omega_2}\mathcal{D}_{\omega}^{(\alpha)}[\rho_{\infty}]\right)\ ,
\end{equation}
whereby the negativity of $J_Q^{(\alpha)}(\rho_\infty)$ denotes increasing heat within the spin chain  and thus heat flowing into it.
In the steady state the total heat flow must vanish, $J_Q(\rho_\infty)=J^{(L)}_Q(\rho_\infty)+J^{(R)}_Q(\rho_\infty)=0$; furthermore, according to the Clausius formulation of the second law of thermodynamics, heat should go from the hotter bath, say the one to the left, into the spin chain and out of it into the colder one at the right end side. Namely,
$\beta_L\leq \beta_R\Longrightarrow J^{(L)}_Q(\rho_\infty)=\,-\,J^{(R)}_Q(\rho_\infty)\leq 0$. However, in~\cite{Levy} it is shown that, for a two-spin chain, this condition holds in the global approach whereas it can be violated in the local one.

In the case of a three spin chain and for the steady state in the global approach given by~\eqref{rog}, the heat flow at the left end,  with
$h_L(\omega_\ell)=h_R(\omega_\ell)=h$, $\ell=0,1,2$, reads 
\begin{align}
\label{eq:HeaetFlow}
J^{(L)}_Q(\rho_\infty)&=\frac{\lambda^2\,h^2\, \pi}{2}\sum_{\ell=0}^2\,\frac{(1+\delta_{\ell,0})\omega_\ell }{\sum_{\alpha=L,R}(2n_{\alpha}(\omega_\ell)+1)}\cr
&
\times\ \Big(n_{R}(\omega_\ell)-n_L(\omega_\ell)\Big)\ .
\end{align}


Since we assumed $\beta_L\leq \beta_R$, one has $n_{L}(\omega_\ell)\geq n_{R}(\omega_\ell)$ for all $\ell=0,1,2$, and heat flows from the left bath into the spin chain and out of it into the right bath, in agreement with Clausius version of the second law of thermodynamics.

Notice that $J^{(L)}_Q(\rho_\infty)$ does not vanish when $g\to0$; instead, in the local approach, one finds $J^{(L)}_Q(\rho_\infty)=0$ at order $g$, namely by 
computing~\eqref{eq:HeatCurrent} with respect to the local stationary state $\rho_\infty=\rho^{(0)}\,+\,g\rho^{(1)}$ in~\eqref{local-steady-state}. In~\cite{Trushechkin}, it is purported that the violation of Clausius second law of thermodynamics in the local approach as observed in~\cite{Levy} can be cured by first observing that the violation occurs at a certain order in the coupling constant $g$ and that the local master equation is the zeroth-order term in that expansion, and secondly, by retaining contributions to the expansion that are of the same order of the violations of the Clausius second law. 
Together with the fact that, for three spins,  the local steady state is not obtainable from the global one by letting $g\to0$, also the behaviour of the global heat flow 
$J^{(L)}_Q(\rho_\infty)$ at $g=0$  seems in contradiction with~\cite{Trushechkin}. However, as already noted there, the extraction of the local approach from the global one at vanishing $g$  is only possible when the degeneracy of the non-interacting spin Hamiltonian is not changing. This is not the case for 
the three spin chain object of the present study, which can indeed also be taken as an evidence of the non-interchangeability of the weak 
coupling limit with the switching off of the inter-spin interaction.

In conclusion, we showed by a fully analytic treatment of a three spin chain  coupled to two heat baths that new physical effects appear, namely the presence in the global approach of sink and source contributions to the time derivative of the spin average which cannot be captured neither in the local approach nor in the case of a simpler two spin chain. Further investigations will be dealing with the extension of our treatment to longer spin chains.

\appendix

\section{}
\label{appendix-a}
In this Appendix we collect the explicit expressions of the eight eigenvalues
and eigenvectors of the system Hamiltonian $H_S$ in (\ref{spin-hamiltonian}).
For the energy levels one gets:
\begin{align}
\nonumber
&E_{1,2}=\pm3\Delta\ ,\quad\quad E_{3,4}=\pm\Delta\ ,\\
&E_{5,6}=\pm\Big(\Delta+2g\sqrt{2}\Big)\ ,\quad E_{7,8}=\pm \Big(-\Delta+2g\sqrt{2}\Big)\ .
\end{align}
The corresponding eigenstates, written in the `computational basis' of tensor product spin states,
$|jk\ell\rangle\equiv |j\rangle\otimes|k\rangle\otimes|\ell\rangle$ with the
convention $\sigma_z | i\rangle =(-1)^i | i\rangle$, $i=0,1$, explicitly read:
\begin{align}
    &\ket{E_{1}}=\ket{000}\\
    &\ket{E_{2}}=\ket{111}\\
    &\ket{E_{3}}=\frac{1}{\sqrt{2}}\big(\ket{001} - \ket{100}\big)\\
    &\ket{E_{4}}=\frac{1}{\sqrt{2}}\big( \ket{011}-\ket{110}\big)\\
    &\ket{E_{5}}=\frac{1}{2}\left(\ket{001}+\sqrt{2}\ket{010}+\ket{100}\right)\\
    &\ket{E_{6}}=\frac{1}{2}\left(\ket{011}-\sqrt{2}\ket{101}+\ket{110}\right)\\
    &\ket{E_{7}}=\frac{1}{2}\left(\ket{011}+\sqrt{2}\ket{101}+\ket{110}\right)\\
    &\ket{E_{8}}=\frac{1}{2}\left(\ket{001}-\sqrt{2}\ket{010}+\ket{100}\right)\ .
\end{align}
In addition, when writing system states as $8\times 8$ density matrices,
we shall use the natural tensor product ordering, namely:
$\ket{000}$, $\ket{001}$, $\ket{010}$, $\ket{011}$, $\ket{100}$, $\ket{101}$, $\ket{110}$, $\ket{111}$.

\section{}
\label{appendix-b}
In this Appendix, we provide additional information on the determination of the stationary state of the dynamics
obtained in the global approach.

The proof of the uniqueness of the stationary state requires re-expressing the Lindblad operators
listed in (\ref{lindblad-operators}) in the basis of the energy eigenstates given in the previous Appendix.
One easily finds:
\begin{align}\label{Eq:LingGlobEn}
    &A_L(\omega_0)=\frac{1}{\sqrt{2}}\big(-\ket{E_3}\bra{E_1}+\ket{E_2}\bra{E_4}\cr
    &\hskip 4cm +\ket{E_7}\bra{E_5}-\ket{E_6}\bra{E_8}\big)\cr
    &A_L(\omega_1)=\frac{1}{2}(\ket{E_8}\bra{E_1}-\ket{E_6}\bra{E_3}-\ket{E_4}\bra{E_5}+\ket{E_2}\bra{E_7})\cr
    &A_L(\omega_2)=\frac{1}{2}(\ket{E_5}\bra{E_1}+\ket{E_7}\bra{E_3}+\ket{E_2}\bra{E_6}+\ket{E_4}\bra{E_8})\cr
    &A_R(\omega_0)=\frac{1}{\sqrt{2}}\big(\ket{E_3}\bra{E_1}-\ket{E_2}\bra{E_4}\cr
    &\hskip 4cm +\ket{E_7}\bra{E_5}-\ket{E_6}\bra{E_8}\big)\cr
    &A_R(\omega_1)=\frac{1}{2}(\ket{E_8}\bra{E_1}+\ket{E_6}\bra{E_3}+\ket{E_4}\bra{E_5}+\ket{E_2}\bra{E_7})\cr
    &A_R(\omega_2)=\frac{1}{2}(\ket{E_5}\bra{E_1}-\ket{E_7}\bra{E_3}+\ket{E_2}\bra{E_6}-\ket{E_4}\bra{E_8})\cr
\end{align}

On the other hand, for the explicit derivation of the stationary state, a diagonal {\it ansatz} in the
spin energy basis suffices:
\begin{equation}
\rho_\infty=\sum_{k=1}^8 \mu_k | E_k\rangle\langle E_k |\ ,
\label{b1}
\end{equation}
The eight unknown constants $\mu_k$ are determined by imposing $\rho_\infty$ to be in the kernel
of the dissipator $\cal D$ in (\ref{global-dissipation-1}). Inserting the expression (\ref{b1})
into the stationary condition ${\cal D}[\rho_\infty]=\,0$ leads to a set of linear equations
that can be represented as
\begin{equation}
{\cal M}\cdot \vec\mu =\,0\ ,
\label{b2}
\end{equation}
where $\vec\mu$ is a 8-dimensional vector with components $\mu_k$, while $\cal M$ is an $8\times8$ matrix
with entries:
\begin{equation}
\mathcal{M}_{k,\ell}=\sum_{\alpha=L,R}\sum_\omega 
\text{Tr}\Big[\mathcal{D}^{(\alpha)}_\omega\big[| E_\ell\rangle\langle E_\ell |\big]|\, E_k\rangle\langle E_k |\Big]\ .
\end{equation}
It can be explicitly expressed in terms of the six quantities reported in~\eqref{tom}-~\eqref{som}. Setting $\tau_i:=\tau(\omega_i)$ and
$s_i:=s(\omega_i)$, $i=0,1,2$, it reads (only the nonvanishing entries are explicitly shown):
%
    \begin{align}
    &\mathcal{M}=\begin{pmatrix}
    m_1 & \cdot & \tau_0 & \cdot & {\tau_1}/{2} & \cdot &\cdot& {\tau_1}/{2}\cr
    \cdot &m_2 & \cdot & s_0 & \cdot & {s_2}/{2} & {s_1}/{2} & \cdot\cr
    s_0 & \cdot & m_3 & \cdot & \cdot & {\tau_1}/{2} & {\tau_2}/{2}& \cdot\cr
    \cdot & \tau_0 & \cdot & m_4 & {s_1}/{2} & \cdot & \cdot & {s_2}/{2}\cr
    {s_2}/{2} & \cdot & \cdot & {\tau_1}/{2} & m_5 & \cdot & \tau_0 & \cdot \cr
    \cdot & {\tau_2}/{2} & {s_1}/{2} & \cdot & \cdot &m_6 & \cdot & s_0\cr
    \cdot & {\tau_1}/{2} & {s_2}/{2} & \cdot & s_0 & \cdot & m_7 & \cdot\cr
    {s_1}/{2} & \cdot & \cdot & {\tau_2}/{2} & \cdot & \tau_0 & \cdot & m_8
    \end{pmatrix}\ ,
\end{align}
%
with the diagonal terms given by
\begin{align}
\label{bmu}
&m_1=-\Big(s_0+\frac{s_1+s_2}{2}\Big)\ ,\quad
m_2=-\Big(\tau_0+\frac{\tau_1+\tau_2}{2}\Big)\ ,\cr
&m_3=-\Big(\tau_0+\frac{s_1+s_2}{2}\Big)\ ,\quad
m_4=-\Big(s_0+\frac{\tau_1+\tau_2}{2}\Big)\ ,\cr
&m_5=-\Big(s_0+\frac{s_1+\tau_2}{2}\Big)\ ,\quad
m_6=-\Big(\tau_0+\frac{\tau_1+s_2}{2}\Big)\ ,\cr
&m_7=-\Big(\tau_0+\frac{s_1+\tau_2}{2}\Big)\ ,\quad
m_8=-\Big(s_0+\frac{\tau_1+\tau_2}{2}\Big)\ .\cr
\end{align}

Together with normalization,  ${\rm Tr}[\rho_\infty]=\sum_k\mu_k=1$, equation (\ref{b2}) uniquely fixes
the components of $\vec \mu$,
\begin{equation}
\vec\mu=\frac{1}{(s_0+\tau_0)(s_1+\tau_1)(s_2+\tau_2)}\begin{pmatrix}
    \tau_0\tau_1\tau_2 \cr
    s_0s_1s_2 \cr
    s_0\tau_1\tau_2 \cr 
    \tau_0s_1s_2 \cr 
    \tau_0\tau_1s_2 \cr
    s_0s_1\tau_2 \cr 
    s_0\tau_1s_2 \cr 
    \tau_0s_1\tau_2
\end{pmatrix}
\end{equation}
and hence the expression of the steady state. Notice indeed that, from~\eqref{tom}-~\eqref{som}, the quantities $\tau_i$ and
$s_i$ are all positive, whence $\rho_\infty$ is a positive and normalized $8\times 8$ matrix.

\section{}
\label{appendix-c}
In this Appendix we shall discuss some general questions regarding the determination in perturbation theory 
of the steady state of a
quantum dynamical semigroup, {\it i.e.} the dynamics generated by a master equation in
Groini-Kossakowski-Sudarshan-Lindblad form.

\subsection{General setting}
\noindent
Let $\gamma_t= e^{t{\cal L}}$ the one-parameter semigroup generated by a master equation as in (\ref{master-equation}),
\begin{equation}
\label{master-equation-a}
{\partial\rho(t)\over \partial t}={\cal L}[\rho(t)]\ ,
\end{equation}
acting on the state space ${\cal S}_d$ of the system, that we assume to be $d$-dimensional,
and denote by $\tilde\gamma_t$ the corresponding `dual' semigroup acting on the system observables (see the discussion leading to 
equation~\eqref{dual-master-equation}),
$\tilde\gamma_t: M_d\mapsto M_d$, where $M_d$ is the set of $d\times d$ complex matrices.
Further, define $\mathcal{G}$ to be the linear map from $\mathcal{S}_d$ into itself constructed through the time-average 
\begin{equation}
\label{EA0}
\mathcal{G}:\mathcal{S}_d\ni\rho\mapsto\mathcal{G}[\rho]=\lim_{T\mapsto+\infty}\frac{1}{T}\int_0^T{\rm d}t\,\gamma_t[\rho]\in\mathcal{S}_d\ .
\end{equation}
Note that $\mathcal{G}$ projects onto the 
stationary manifold of $\gamma_t$:
\begin{equation}
\label{EA1}
\mathcal{G}\circ\gamma_t=\gamma_t\circ\mathcal{G}=\mathcal{G}=\mathcal{G}^2
\ ,\ \gamma_t[\rho]=\rho\Longleftrightarrow \mathcal{G}[\rho]=\rho\ ,
\end{equation}
or equivalently in terms of the generator:
\begin{equation}
\label{EA2}
\mathcal{G}\circ\mathcal{L}=\mathcal{L}\circ\mathcal{G}=0\ ,\ \mathcal{L}[\rho]=0\Longleftrightarrow \mathcal{G}[\rho]=\rho\ .
\end{equation}

Let us first discuss the conditions for the inverse  $\mathcal{L}^{-1}$ of the generator $\mathcal{L}$ to exist.
Clearly, it can be well defined only on a subspace that does not contain elements of the 
kernel of $\mathcal{L}$. To this purpose, consider the operator $\mathcal{F}:={\rm id}-\mathcal{G}$. Notice that
\begin{equation}
\label{EA3}
\mathcal{L}\circ\mathcal{F}=\mathcal{F}\circ\mathcal{L}\ ,
\end{equation}
so that the range of $\mathcal{F}$, ${\rm Ran}(\mathcal{F})$, is mapped into itself by  the generator $\mathcal{L}$.
Extending  $\mathcal{G}$ by linearity on the whole of $M_d$, one obtains that, if $\mathcal{G}[X]=0$ for $X\in M_d$, 
then automatically $X\in{\rm Ran}(\mathcal{F})$ as
\begin{equation}
\label{EA3a}
\mathcal{G}[X]=0\Longrightarrow X=X-\mathcal{G}[X]=\mathcal{F}[X]\ .
\end{equation}
Thus ${\rm Ker}({\cal G})\subseteq {\rm Ran}({\cal F})$. Moreover, 
\begin{equation}
\label{EA4}{\rm Ran}(\mathcal{F})\cap{\rm Ker}(\mathcal{L})=0\ .
\end{equation}
Indeed, if $X=Y-\mathcal{G}[Y]$ and $\mathcal{L}[X]=0$, from~\eqref{EA2} it follows that
\begin{align}
\nonumber
\mathcal{L}[X]&=0=\mathcal{L}[Y]-\mathcal{L}\circ\mathcal{G}[Y]\Longrightarrow\mathcal{L}[Y]=0\\
&\Longrightarrow \mathcal{G}[Y]=Y\Longrightarrow X=Y-Y=0\ .
\label{EA5}
\end{align}
The inverse $\mathcal{L}^{-1}$ can then be defined as the map from 
${\rm Ran}(\mathcal{F})$ into ${\rm Ran}(\mathcal{F})$ such that
\begin{equation}
\mathcal{L}\circ\mathcal{L}^{-1}=\mathcal{L}^{-1}\circ\mathcal{L}\ .
\label{EA6}
\end{equation}
As inverse of $\mathcal{L}$ on ${\rm Ran}(\mathcal{F})$, $\mathcal{L}^{-1}$ satisfies 
\begin{equation}
\label{EA6a}
\mathcal{L}^{-1}\circ\mathcal{G}=\mathcal{G}\circ\mathcal{L}^{-1}\ ,
\end{equation} 
whence $\mathcal{L}^{-1}({\rm Ran}(\mathcal{F}))\subseteq{\rm Ran}(\mathcal{F})$ 
and ${\rm Tr}\circ\mathcal{L}^{-1}=0$; indeed, the trace-preserving property of $\gamma_t$, 
and thus of $\mathcal{G}$, entails ${\rm Tr}[X]=0$ for all $X\in{\rm Ran}(\mathcal{F})$.

\subsection{Perturbative expansion}
\noindent
Suppose now that the generator $\cal L$ of $\gamma_t$ has the form as in (\ref{L-expansion}),
\begin{equation}
\mathcal{L}_g=\mathcal{L}_0+g\,\mathcal{L}_1\ ,
\end{equation}
where $g$ is a small perturbative parameter. We seek a perturbative expansion 
of the stationary states \hbox{$\mathcal{L}[\rho_\infty]=0$}, in the form
\begin{equation}
\label{SS0}
\rho_\infty=\sum_{n=0}^\infty g^n\,\rho^{(n)}\ ;
\end{equation}
leading to the following recursive relations:
\begin{equation}
\label{SS1}
\mathcal{L}_0[\rho^{(0)}]=0\ ,\qquad \mathcal{L}_0[\rho^{(n+1)}]=-\mathcal{L}_1[\rho^{(n)}]\ .
\end{equation}
Therefore, once a stationary state $\rho^{(0)}$ of $\mathcal{L}_0$ is chosen, its first order perturbation 
can be obtained by the inversion of $\mathcal{L}_0$:
\begin{equation}
\label{SS2}
\rho^{(1)}=-\mathcal{L}_0^{-1}\circ\mathcal{L}_1[\rho^{(0)}]\ .
\end{equation}
From the previous subsection, we know that this can be done by  ensuring that $\mathcal{L}_1[\rho^{(0)}]\in{\rm Ran}(\mathcal{F})$.
If there are more than one stationary state $\rho^{(0)}$ of $\mathcal{L}_0$, then, according to~\eqref{EA3a}, this property can be enforced
by adjusting $\rho^{(0)}$
so that 
\begin{equation}
\label{SS3}
\mathcal{G}_0\circ\mathcal{L}_1[\rho^{(0)}]=0\ ,
\end{equation}
where $\mathcal{G}_0$ is the average map associated with the semigroup generated by $\mathcal{L}_0$. This 
implies $\mathcal{L}_1[\rho^{(0)}]\in{\rm Ran}(\mathcal{F}_0)$ and thus $\notin{\rm Ker}(\mathcal{F}_0)$, where $\mathcal{F}_0={\rm id}-\mathcal{G}_0$.
This same argument can be applied at all orders since all of them ask for the inversion of $\mathcal{L}_0$.

\subsection{Application to the $XX$ spin-chain}
\noindent
The application of the previous general considerations to the specific case discussed in
the main text is straightforward. Following the definitions and conventions of Section \ref{sec:local},
one first realizes that 
the average map $\mathcal{G}_0$ with respect to $\displaystyle\gamma^{(0)}_t={\rm e}^{t\,\mathcal{L}_0}$ 
can be cast in the form:
\begin{equation}
\label{3q7}
\mathcal{G}_0[\rho]=\lambda_+(\rho)\,\rho_+\,+\,\lambda_-(\rho)\,\rho_-\ ,
\end{equation}
where $\lambda_\pm(\rho)\geq 0$, $\lambda_+(\rho)+\lambda_-(\rho)=1$ and
\begin{equation}
\label{3q8}
\rho_\pm=\rho_L\otimes\frac{{\bf 1}\pm\sigma_z}{2}\otimes\rho_R\ .
\end{equation}
Now, the operators $P_\pm$  in~\eqref{p-operators} are such that:
\begin{equation}
\label{3q10}
P_\pm\,\rho_\pm\,=\,\rho_\pm\,P_\pm\,=\,\rho_\pm\ ,\qquad P_\pm\,\rho_\mp\,=\,0\ .
\end{equation}
Moreover, they are left invariant by the dual semigroup (see~\eqref{dual-master-equation} and the discussion preceding it); 
indeed, the dual of the dissipative part and of the Hamiltonian contribution of the corresponding generator
$\widetilde {\cal L}_0$ are such that 
\begin{equation}
\label{3q10a}
\widetilde{\mathcal{D}}_{L,R}[{\bf 1}]=0\ ,\qquad \Big[\Delta\sum_{i=1}^3\sigma_3^{(i)}\,,\,P_\pm\Big]=0\ ,
\end{equation}
so that, $\widetilde{\mathcal{L}}_0[P_\pm]=0$, whence
$\widetilde{\mathcal{G}}_0[P_\pm]=P_\pm$ under the dual $\widetilde{\mathcal{G}}_0$ of the average map $\mathcal{G}_0$.
Then, using~\eqref{3q7}
\begin{eqnarray}
\nonumber
&&\hskip -.5cm {\rm Tr}\Big[P_\pm\,\mathcal{G}_0[\rho]\Big]={\rm Tr}\Big[\widetilde{\mathcal{G}}_0[P_\pm]\,\rho\Big]
={\rm Tr}\Big[P_\pm\,\rho\Big]\\
&&=\lambda_+(\rho)\,{\rm Tr}\Big[P_\pm\,\rho_+\Big]\,+\,\lambda_-(\rho)\,{\rm Tr}\Big[P_\pm\,\rho_-\Big]\ ,
\label{3q11}
\end{eqnarray}
yields $\lambda_\pm(\rho)={\rm Tr}\big[P_\pm\,\rho\big]$, and thus
\begin{equation}
\label{3q12}
\mathcal{G}_0[\rho]=\rho_L\otimes\begin{pmatrix}{\rm Tr}\Big[\rho\,P_+\Big]&0\cr
0&{\rm Tr}\Big[\rho\,P_-\Big]\end{pmatrix}\otimes\rho_R\ .
\end{equation}
In order to obtain the $n$-th order perturbation,
\begin{equation}
\label{3q14}
\rho^{(n)}=-\mathcal{L}_0^{-1}\circ\mathcal{L}_1[\rho^{(n-1}]\ ,
\end{equation}
one now needs to invert $\mathcal{L}_0$.
According to the general construction developed above, in order to do that we first proceed to ensure that
\begin{equation}
\mathcal{G}\circ\mathcal{L}_1[\rho^{(n-1)}]=\rho_L\otimes
\begin{pmatrix}
\mit\Gamma_+^{(n-1)} &0\cr
0&\mit\Gamma_-^{(n-1)}
\end{pmatrix}
\otimes\rho_R=\,0\ ,
\label{3q15}
\end{equation}
where
\begin{equation}
{\mit\Gamma}_\pm^{(n-1)}={\rm Tr}\Big[\mathcal{L}_1[\rho^{(n-1)}]\,P_\pm\Big]\ .
\end{equation}
This request together with $P_+\,+\,P_-={\bf 1}$ and the trace-preserving character of $\mathcal{L}_1$ implies that one may need to adjust $\rho^{(n-1)}$ so that
one of the following two equivalent conditions holds true:
\begin{equation}
\label{3q16}
{\mit\Gamma}_+^{(n-1)}=-{\mit\Gamma}_-^{(n-1)}=0\ .
\end{equation}
In the case $n=2$, this is precisely the condition fixing uniquely the value of the parameter
$r$ given in (\ref{value-r}).
\vskip 1cm

\section*{acknowledgments}

L. M. acknowledges financial support by Sharif University
of Technology, Office of Vice President for Research
under Grant No. G930209 and hospitality by the Abdus
Salam International Centre for Theoretical Physics
(ICTP) where parts of this work were completed. F. B. and R. F. 
acknowledge that their research has been conducted within the framework of the Trieste Institute for Theoretical Quantum Technologies.

The authors contributed equally to this study and are listed in alphabetical order.


\end{document}